\documentclass{jfm}
\usepackage{graphicx}
\usepackage{epstopdf, epsfig}

\usepackage[intlimits]{amsmath}

\shorttitle{Erosion and deposition in an internally convective fluid.}
\shortauthor{C. Sturtz, \'E. Kaminski, A. Limare and S. Tait.}

\title{The fate of particles in a volumetrically heated convective fluid at high Prandtl number.}

\author{Cyril Sturtz\aff{1}
  \corresp{\email{sturtz@ipgp.fr}},
 \'Edouard Kaminski\aff{1},
 Angela Limare\aff{1}
 \and Stephen Tait\aff{1}}

\affiliation{\aff{1}Universit\'e de Paris, Institut de Physique du Globe de Paris, CNRS, F-75005 Paris, France.
}

\begin{document}

\maketitle

\begin{abstract}
\indent The dynamics of suspensions plays a crucial role on the evolution of geophysical systems such as  lava lakes, magma chambers and magma oceans. During their cooling and solidification, these magmatic bodies involve convective viscous fluids and dispersed solid crystals that can form either a cumulate or a floating lid by sedimentation. We study such systems based on internal heating convection experiments in high Prandtl fluids bearing plastic beads. We aim to determine the conditions required to produce a floating lid or a sedimented deposit. We show that although the sign of particles buoyancy is the key parameter, it is not sufficient to predict the particles fate. To complement the model we introduce the Shields formalism and couple it with scaling laws describing convection. We propose a generalized Shields number that enables a self-consistent description of the fate of particles in the system, especially the possibility to segregate from the convective bulk. We provide a quantification of the partition of the mass of particles in the different potential reservoirs (bulk suspension, floating lid, settled cumulate) through reconciling the suspension stability framework with the Shields formalism. We illustrate the geophysical implications of the model by revisiting the problem of the stability of flotation crusts on solidifying rocky bodies.
\end{abstract}

\begin{keywords}
Erosion, suspension, convection, internal heating, magma ocean, cumulate, flotation crust, secular cooling.
\end{keywords}

\setcounter{tocdepth}{3}

\section{Introduction}
\indent According to the classical scenarii of planetary formation, terrestrial bodies were likely  partially or totally molten, forming a magma ocean \citep{Taylor92, Tonks93,Abe95,Abe97}. This initial stage in planetary history is due to two major phenomena. In the first few million years of the solar system, during the accretion of planetesimals, the decay of short lived radioactive elements such as $^{26}\rm{Al}$ and $^{60}\rm{Fe}$ is an important heating source \citep{Urey55, Neumann12,Weidenschilling19,Kaminski20}. Later, collisions between planetary embryos and giant impacts converted gravitational energy into heat and produced massive melting events \citep{Tonks92b,Safronov94,Canup01}. During the cooling of such a system, the temperature evolves from the liquidus to the solidus, a melting interval which is several hundred of degrees for silicate systems. If the cooling rate controls the pace of solidification of the system, solidification and the fate of crystals can also introduce a feedback on the thermal history of the system. For example, the anorthosite crust of the Moon formed by flotation of light plagioclase crystals \citep{Wood72,Warren85,Shearer06}. This thick crust has an insulating effect on the convective system underneath \citep{Lenardic03,Grigne07} hence slowed down its thermal evolution \citep{ElkinsTanton11,Maurice20}.\\
\indent  If the fluid were quiescent, crystals would settle down or float according to the sign of their buoyancy. Convection may prevent this behavior by maintaining particles in suspension. The crucial issue lies in the determination of the stability of such a suspension. \cite{Sparks84} pointed out that the Stokes velocity of particles in magma chambers, which stands for the typical settling velocity, is small compared to the mid-depth vertical root-mean square fluid velocity. In that case, suspensions in magmatic reservoirs should always be stable, as particles behave like passive tracers. However, \cite{MartinNokes88} illustrated experimentally that negatively buoyant particles initially in suspension  eventually settle down and form deposits. \cite{Lavorel09} furthermore highlighted that deposition occurs at Stokes velocity even though convection is turbulent. These observations can be interpreted considering the interaction between particles and the dynamical boundary layers that develop at the borders of the reservoir, where velocities vanish because of rigid boundary conditions \citep{Sparks84}. As a matter of fact, dealing with suspension sustainability requires a criterion that involves both convection and sedimentation. \cite{Solomatov93b} proposed a description based on the energy balance of fluid-particle interaction. These authors assumed that the fluid can transfer an amount $\epsilon$ of its convective energy to particles. If the suspension gravitational energy that drives settling exceeds this quantity, the suspension is not stable and deposits form. Even though $\epsilon=0.1-1\ \%$ had been evaluated from experiments \citep{Solomatov93a,Lavorel09}, it has been underlined that this ad-hoc parameter is not well constrained \citep{Solomatov93a}. If it were, this model could make possible a totally self-consistent mass budget, without involving any ad-hoc parametrization.\\
\indent Instead of dealing with suspension stability, other authors prefer to consider the stability of the beds of particles that may form. Erosion and entrainment of particles is usually described by the formalism proposed by \cite{Shields36}. Basically, particles are locked on the bed surface because of frictional forces that have the same order of magnitude as the buoyancy force according to Coulomb's law. Entrainment is possible if the ratio of the stress acting on particles over the particles buoyancy exceeds a critical value that lies between $0.1-0.2$ \citep{Charru04}. This model was used to determine sediment transport upon bedloads \citep{Lajeunesse10} and the equilibrium height of the settled bed \citep{Leighton86}. However, these studies refer to experiments that involve controlled flow, with isoviscous and isothermal conditions, that are not consistent with the geophysical applications considered here. Convection in magmatic reservoirs involves destabilization of thermal boundary layers (TBL) that complexifies both temperature and velocity fields. \cite{Solomatov93a} adapted the previous reasoning by considering that a single particle in the TBL can not be lifted, but is moved horizontally at the surface of the bed by the tangential stress. Then, particles form dunes that make the stress quasi-vertical, enabling entrainment. However, these authors do not study the equilibrium thickness of the underlying bed, or its influence on the thermal state. \\
\indent In the present study, we aim at quantifying the behavior of such a lid embedded within the TBL of a convective system. We will consider the case of an internally heated system cooled from above, displaying only one TBL which is at the upper surface of the convective layer, and we will focus on the stability and the equilibrium thickness of a floating lid. The first part of the study tackles the dimensionless equations that outline the problem, and underlines the key dimensionless numbers that describe the system. In the second part, we develop an experimental approach to study this issue. The setup used is composed of a tank containing the convective fluid internally heated by microwave absorption and plastic beads that represent crystals. We examine the erosion of the floating lid, and we introduce a model that predicts the equilibrium thickness of the crust according to the thermal state. We identify and test experimentally the stability condition to form a cumulate. We emphasize one dimensionless parameter, the Shields number, as the key parameter to deal both with the lid stability and with the suspension sustainability. Finally, we discuss a geological case illustrating the relevance of our model: the condition of stability of a floating lid on terrestrial bodies.


\section{Theoretical framework}


	\subsection{Internally heated convective systems}	
	Geophysical problems considered here involve convective systems driven by internal heating and secular cooling, phenomena that are mathematically equivalent. Thermal convection in the Boussinesq approximation is governed by the following equations representing the conservation of momentum, energy and mass \citep[see, e.g.:][pp.114]{JaupartMareschal}:
	\begin{eqnarray}
		\rho_{0,f} \left(\frac{\partial \mathbf{u}_{f}}{\partial t}+\mathbf{u}_{f} \cdot \nabla \mathbf{u}_{f}\right) &=& - \nabla P + \eta_{f} \nabla^{2} \mathbf{u}_{f} - \alpha_{f} \rho_{0,f} \theta \mathbf{g}, \label{eq:21}\\
		\rho_{0,f} c_{p,f} \left(\frac{\partial T}{\partial t}+ \mathbf{u}_{f}\cdot \nabla T\right)&=&\lambda_{f} \nabla^{2} T + H, \label{eq:22}\\
		\nabla \cdot \mathbf{u}_{f}&=&0, \label{eq:23}
	\end{eqnarray}
	where $\mathbf{u}_{f}$ is the fluid velocity field, $\rho_{0,f}$ is the reference fluid density, $\eta_{f}$ is its dynamic viscosity, $\alpha_{f}$ is its thermal expansion, $c_{p,f}$ is its specific heat, $\lambda_{f}$ is its thermal conductivity,  $\mathbf{g}$ is the acceleration of gravity, $H$ is the rate of internal heat generation, $P$ is the pressure, $T$ is the temperature field, and $\theta=T-\langle T\rangle$ is the thermal anomaly with $\langle T\rangle$  the horizontal average temperature.\\
	\indent Internally heated convective systems are characterized by the internal temperature scale $\Delta T_{H}$ defined as:
	\begin{equation}
		\Delta T_{H}=\frac{Hh^{2}}{\lambda_{f}}, \label{eq:24}
	\end{equation}
	where $h$ is the vertical thickness of the system. This temperature scale introduces a new definition of the Rayleigh number, called the Rayleigh-Roberts number \citep{Roberts67}:
	\begin{equation}
		Ra_{H}=\frac{\alpha_{f} \rho_{0,f} g Hh^{5}}{\eta_{f} \kappa_{f} \lambda_{f}}, \label{eq:25}
	\end{equation}
	 where $\kappa_{f}=\lambda_{f}/\rho_{0,f}c_{p,f}$ is the fluid's thermal diffusivity. This dimensionless number enables the characterization of the vigor and patterns of convection \citep[e.g.:][]{Vilella18}.\\
	\indent Using $h$ as the length scale, $\Delta T_{H}$ as the temperature scale, $W=\rho_{0,f}\alpha_{f}g\Delta T_{H} h^{2}/\eta_{f}$ as the velocity scale that represents the Stokes' velocity of a laminar thermal, $h/W$ as the time scale, and $\eta_{f}W/h$ as the pressure scale, one can provide dimensionless form of the Boussinesq equations as follows \citep[see, e.g.:][pp.114]{JaupartMareschal}:
	\begin{eqnarray}
		Ra_{H} Pr^{-1}\left(\frac{\partial \mathbf{u}_{f}}{\partial t}+\mathbf{u}_{f} \cdot \nabla \mathbf{u}_{f}\right) &=& - \nabla P +  \nabla^{2} \mathbf{u}_{f} - \theta \mathbf{e_{z}}, \label{eq:26}\\
		Ra_{H} \left(\frac{\partial T}{\partial t}+ \mathbf{u}_{f}\cdot \nabla T\right)&=&\nabla^{2} T + 1,\label{eq:27}\\
		\nabla \cdot \mathbf{u}_{f}&=&0, \label{eq:28}
	\end{eqnarray}
	where $Pr$ is the Prandtl number, which compares viscous and thermal diffusion:
		\begin{equation}
			Pr=\frac{\nu_{f}}{\kappa_{f}}, \label{eq:29}
		\end{equation}		
		where $\nu_{f}=\eta_{f}/\rho_{f}$ is the kinematic viscosity. The Prandtl and Rayleigh-Roberts numbers characterize the regime of convection occurring in the system. They scale inertia in equation (\ref{eq:26}). The high-$Pr$ low-$Ra_{H}$ limit corresponds to laminar flows, whereas low-$Pr$  high-$Ra_{H}$ yields turbulent inertial flows. In geophysical systems, these dimensionless numbers span a large range of values. For instance, solid-state convection in the current Earth's mantle verifies $Pr\approx 10^{23}$ and $Ra\approx 10^{7}$, whereas magmatic reservoirs are rather evolving with $Pr\approx 10^{3}-10^{8}$ and $Ra\approx 10^{11}-10^{16}$. In the case of magma oceans, huge variation of $Pr$ and $Ra_{H}$ over the thermal history are expected. \cite{Massol16} estimated for the terrestrial magma ocean a Rayleigh number that goes from $10^{31}$ at the very beginning if the planet is totally molten, to $10^{14}$ when crystals are in suspension. This drop can partly be explained by the fact that the presence of crystals increases the apparent viscosity of the mixture by several orders of magnitude when the rheological transition is reached \citep{Lejeune95,Guazzelli18}. As a consequence, the $Pr$ number increases drastically during the solidification, from an initial value around $10^{1}-10^{2}$ to the current value of $10^{23}$ for solid mantle convection. In comparison, experiments carried out in the present study lie in the following ranges: $Pr\approx 10^{3}$ and $Ra_{H}\in [3.10^{6},\ 10^{8}]$. According to the theory by \cite{Grossmann00,Grossmann01}, partially crystallized magma oceans and our experiments occur in the same regime of convection.\\
	\indent Internally heated convective systems are characterized by a single upper TBL generating cold instabilities. The temperature drop across the TBL $\Delta T_{TBL}$ and the velocity of downwellings $W_{i}$ have been studied experimentally and numerically, and can be expressed based on local scaling analyses \citep{Limare15,Vilella18}:
\begin{eqnarray}
	\Delta T_{TBL}&=&C_{T}\Delta T_{H} Ra_{H}^{-1/4}, \label{eq:210}\\
	W_{i}&=&C_{W}\frac{\kappa_{f}}{h} Ra_{H}^{3/8},  \label{eq:211}
\end{eqnarray}
	where the pre-factors $C_{T}$ and $C_{W}$ depend only on the mechanical boundary condition at the top of the system (see: table \ref{tab:coeff}).

\begin{table}
\centering
	\begin{tabular}{ c c c }
		\hline
		Conditions & $C_{T}$ & $C_{W}$\\
		\hline
		 Rigid & 3.41 &  0.732\\
		 Free-slip & 2.49 & 1.216 \\
		\hline
	\end{tabular}
	\caption{Pre-factors of scaling laws (\ref{eq:210}) and (\ref{eq:211}) determined numerically \citep{Vilella18}.}
	\label{tab:coeff}
\end{table}

	
		Introducing particles into the convective fluid add significant complexity in the problem. Indeed, depending on the density, shape and size of the particles, but also depending on the fluid flow, multiple phenomena are likely to occur, which are described in an abundant literature \citep{Andreotti11, Boyer11, Houssais15}.  We summarize below the theoretical framework and fundamental parameters that will later be used to model these interacted phenomena.

	\subsection{Particles in suspension -- two-phase flow}


	The presence of crystals dispersed in a magma reservoir makes it a two-phase system. To describe the dynamics of two-phase flow, a set of complementary equations has to be added to the conservation equations (\ref{eq:21})$-$(\ref{eq:23}) to describe the two phases and the interactions between them \citep[see, e.g.:][pp.306]{Andreotti11}. Considering the fluid phase ($f$) and the solid phase ($p$), and assuming that the volume fraction $\phi$ of the solid phase is uniform, i.e. no chemical or mass exchanges between the phases, equations of motion for the fluid and the particles are respectively: 
	\begin{eqnarray}
		(1-\phi)\rho_{0,f}\left(\frac{\partial \mathbf{u}_{f}}{\partial t}+\mathbf{u}_{f}\cdot\nabla \mathbf{u}_{f}\right)&=&-(1-\phi)\nabla P+(1-\phi) \eta_{f} \nabla^{2} \mathbf{u}_{f},  \nonumber\\
&& \hspace{3cm}-(1-\phi)\rho_{0,f}\alpha_{f}\theta\mathbf{g}-\mathbf{f},   \label{eq:212}   \\
		\phi \rho_{0,p}\left( \frac{\partial \mathbf{u}_{p}}{\partial t}+\mathbf{u}_{p}\cdot \nabla \mathbf{u}_{p}\right) &=&\phi \Delta \rho \mathbf{g} +\mathbf{f}, \label{eq:213}
	\end{eqnarray}
		where $\Delta \rho=\rho_{f}-\rho_{p}$ is the density difference between the fluid and the particles, and $\mathbf{f}$ is the fluid-particle interaction force. All other forces are neglected, such as Van der Waals interactions or frictional forces between particles. There are many approaches describing the interaction force $\mathbf{f}$ depending on the flow regime and particle properties \citep{Maxey83}. Here, we consider that at high $Pr$ number, the interaction between the fluid and particles is dominated by the viscous drag, which is written as:
		\begin{equation}
			\mathbf{f}=\beta(\phi)\frac{\eta_{f}}{r^{2}}(\mathbf{u}_{f}-\mathbf{u}_{p}), \label{eq:214}
		\end{equation}
		where $r$ is the particle radius and $\beta(\phi)$ is a dimensionless function that refers to the contribution of the other particles to the drag. It increases as $\phi$ increases  \citep[pp.306]{Andreotti11}. \\
		\indent Using the same scales as before, the dimensionless set of equations becomes:
		\begin{eqnarray}
		\frac{\partial \mathbf{u}_{f}}{\partial t}+\mathbf{u}_{f}\cdot\nabla \mathbf{u}_{f}&=&Ra_{H}^{-1}Pr\left(\nabla P+\nabla^{2} \mathbf{u}_{f}-\theta\mathbf{e_{z}}\right)-\frac{\beta(\phi)}{1-\phi}\frac{\rho_{0,p}}{\rho_{0,f}}St^{-1} (\mathbf{u}_{f}-\mathbf{u}_{p}), \label{eq:215}\\
		 \frac{\partial \mathbf{u}_{p}}{\partial t}+\mathbf{u}_{p}\cdot \nabla \mathbf{u}_{p}&=&\mathcal{C}\mathbf{e_{z}} + \frac{\beta(\phi)}{\phi}St^{-1}(\mathbf{u}_{f}-\mathbf{u}_{p}).  \label{eq:216}
		\end{eqnarray}
		Two other dimensionless numbers appear. The $\mathcal{C}$ parameter compares the gravitational potential energy of particles to the inertial drag:
			\begin{equation}
				\mathcal{C}=\frac{\Delta \rho g h}{\rho_{0,p}W^{2}}. \label{eq:217}
			\end{equation}
			In our experiments, this parameter is of order unity. The Stokes number characterizes the interaction between the fluid and particles:
			\begin{equation}
				St=\frac{\rho_{0,p}r^{2}W}{\eta_{f} h}. \label{eq:218}
			\end{equation}
		For reservoirs much larger than the size of particles, which is a limit relevant for magma oceans or magma reservoirs, and/or for laminar flow,  the Stokes number is likely to be much smaller than unity. In this case, particles are statistically passive tracers, following fluid motions \citep[e.g.:][pp.25]{Crowe12}. Consequently, (\ref{eq:216}) becomes:
		\begin{equation}
			 \mathbf{u}_{p}=\mathbf{u}_{f}. \label{eq:219}
		\end{equation}
			 \indent We emphasize that this limit describes the average behavior of particles but does not imply that particles never settle. Over a short timescale compared to the convective timescale, the particles are indeed passive tracers and the equality (\ref{eq:219}) holds true. Nevertheless, particles are still buoyant, so there is a small component of the particles velocity that participates to settling. In this way, equation (\ref{eq:219}) rewrites $\mathbf{u}_{p}=\mathbf{u}_{f} + \mathbf{u}_s$, with $||\mathbf{u}_s||\ll ||\mathbf{u}_f||$. Thus, sedimentation actually occurs with a low probability \citep{Patocka20}, and deposits form at long timescales compared to the convective one.\\
		\indent A direct consequence of particles sedimentation is the formation of settled cumulates or floating lids. This implies in turn that erosion and particles re-entrainment from these layers must be taken into account in the modeling of convective systems.

		
		\subsection{Settling and re-entrainment}
		Erosion and re-entrainment from settled cumulates and/or floating lids bring particles back in suspension \citep{Solomatov93a}. However, the framework that describes this phenomenon is different from the one used to treat suspensions as it depends on local mechanical equilibrium of particles \citep{Charru04,Lajeunesse10}. Particles at the surface of the bed are submitted to two forces: (i) the frictional force between the particles and the underlying bed that captures particles at the surface of the bed and is proportional to the particles buoyancy according to Coulomb's law and (ii) the shear stress induced by the flow. A dimensionless number, called the Shields number, compares these two forces \citep{Shields36}:
		\begin{equation}
			\zeta=\frac{\tau}{\Delta \rho g r}, \label{eq:220}
		\end{equation}
		where $\tau$ is the shear stress at the surface of the bed. This ratio enables the definition of a critical value $\zeta_{c}$ that describes the threshold behavior of particles on the bed. If $\zeta<\zeta_{c}$, particles are locked on the bed by frictional forces, whereas if $\zeta>\zeta_{c}$, the shear stress is strong enough to erode particles. For spherical plastic particles homogeneously sheared by a viscous, laminar flow, \cite{Charru04} estimated $\zeta_{c}=0.12$. \\
		\indent The force driving particles (re-)entrainment does not take into account any vertical pressure effects. This follows the comment made by \cite{Solomatov93a} that a single particle can not be lifted by a vertical pressure gradient by comparing the pressure force exerted on the particle to its buoyancy.  The authors proposed that the mechanism for entrainment of particles is strongly linked to shear stress acting at the interface, that manage to displace particles and form dunes that enables entrainment. \\
		\indent  The framework presented above can be used to describe the dynamics of a suspension and the coupled stability of cumulates and/or floating layers. To our knowledge, this coupling has not been studied yet in internally heated convective systems relevant for magma reservoirs. To fill this gap, we present below lab-scale experiments.
		

\section{Experimental approach}


\begin{figure}
	\centering
	\includegraphics[width=\textwidth]{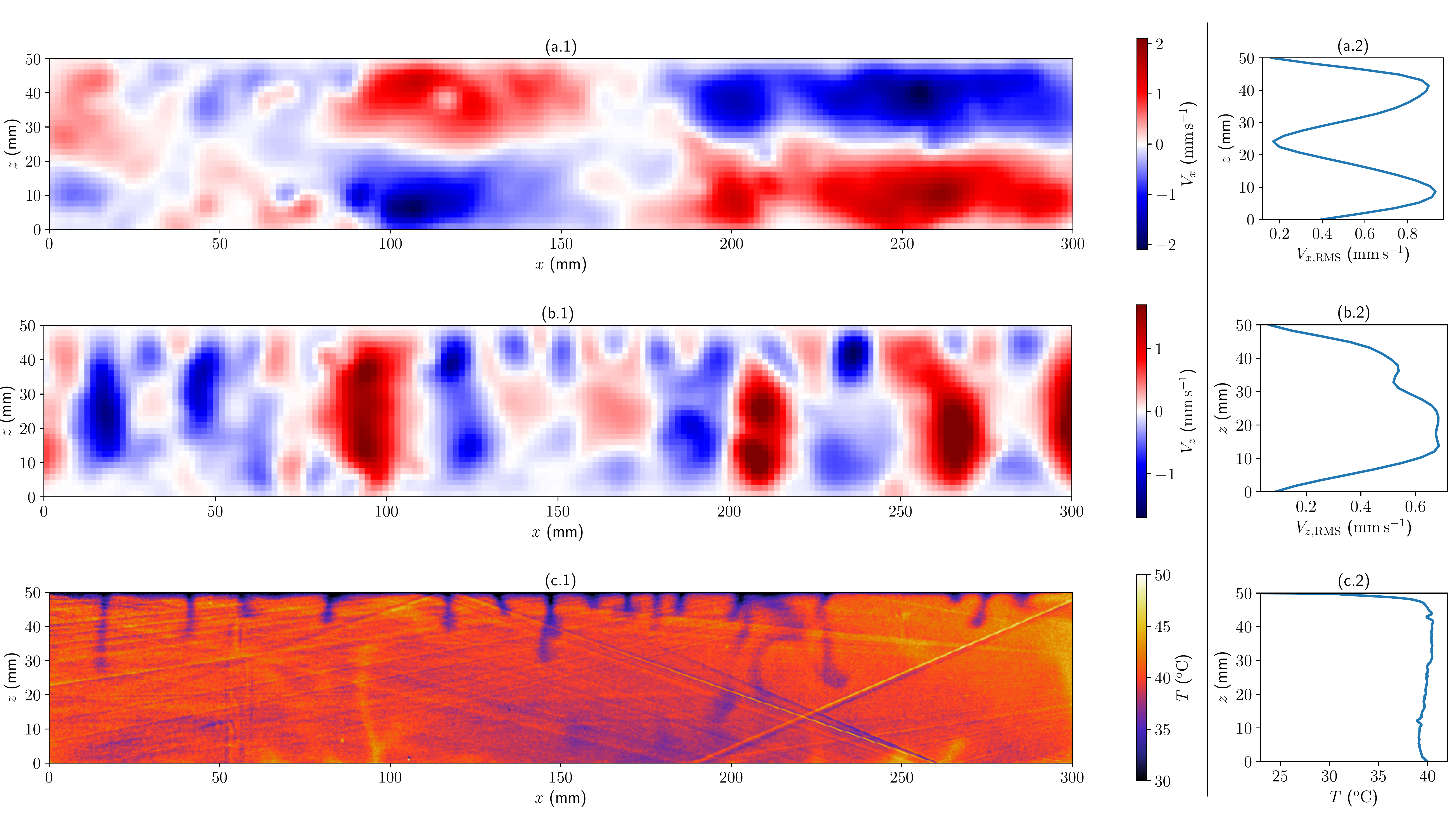}
	\caption{2D horizontal velocity (a.1), vertical velocity (b.1) and temperature (c.1) fields for an experiment without beads ($\rm{IHB29\_3}$). We display the corresponding RMS vertical profiles for the horizontal and vertical velocity in (a.2) and (b.2) respectively, and the average temperature profile (c.2).}
	\label{fig:Exp_field}
\end{figure}

	\subsection{Convection with internal heating}
	
	\indent Achieving experimental convection driven by homogeneous internal heating at high Rayleigh numbers was challenging until \cite{Limare13} developed a unique experimental set-up based on microwave absorption. A $30\times 30\, \mathrm{cm^{2}}$ wide and $5\, \mathrm{cm}$ high tank is introduced in a specially designed microwave oven \citep{Surducan14}. The top of the tank is a thermostated aluminium plate whose temperature is fixed and monitored. The other walls and the base of the tank are made of poly(-methyl methacrylate) (PMMA), transparent to visible light and microwave radiation, and ensuring adiabatic thermal boundary conditions. \\
	\indent A laser sheet scans half of the tank ($15\, \mathrm{cm}$), and we acquire images at a spacing of $1\, \mathrm{cm}$. Two CCD cameras register images in different spectral ranges allowing non-invasive measurement of the temperature field via a two-dye Laser Induced Fluorescence (LIF) method. The velocity field is measured by Particle Image Velocimetry (PIV). The temperature and velocity spatial resolution are $0.2$ and $0.8\, \mathrm{mm}$ respectively. Further details on the experimental setup and calibration can be found in \cite{Fourel17}. The same set-up and methods are used in the following study, but the fluid is adapted to study the sedimentation of beads. Typical 2D velocity  and temperatures field are shown in figure \ref{fig:Exp_field} for experiments without beads. Panels (a.1) and (b.1) show the 2D horizontal and vertical velocity fields and their correspondent root mean square (RMS- vertical profiles (panels ((a.2) and (b.2)). The velocity is zero on the boundaries since they are all rigid. Negative vertical  velocities are associated with thermal instabilities generated at the top boundary of the tank. Positive vertical velocities corresponds only to return flow. Figure \ref{fig:Exp_field} (c.2) displays the temperature vertical profile showing that the thermal structure of the convective layer can be split into an upper boundary layer and a convective interior. An important feature is that the fluid interior has slightly negative temperature gradient. This has been verified numerically \citep{Sotin99,Vilella18} and experimentally \citep{Limare15}.

\begin{figure}
	\centering
	\includegraphics[width=0.5\textwidth]{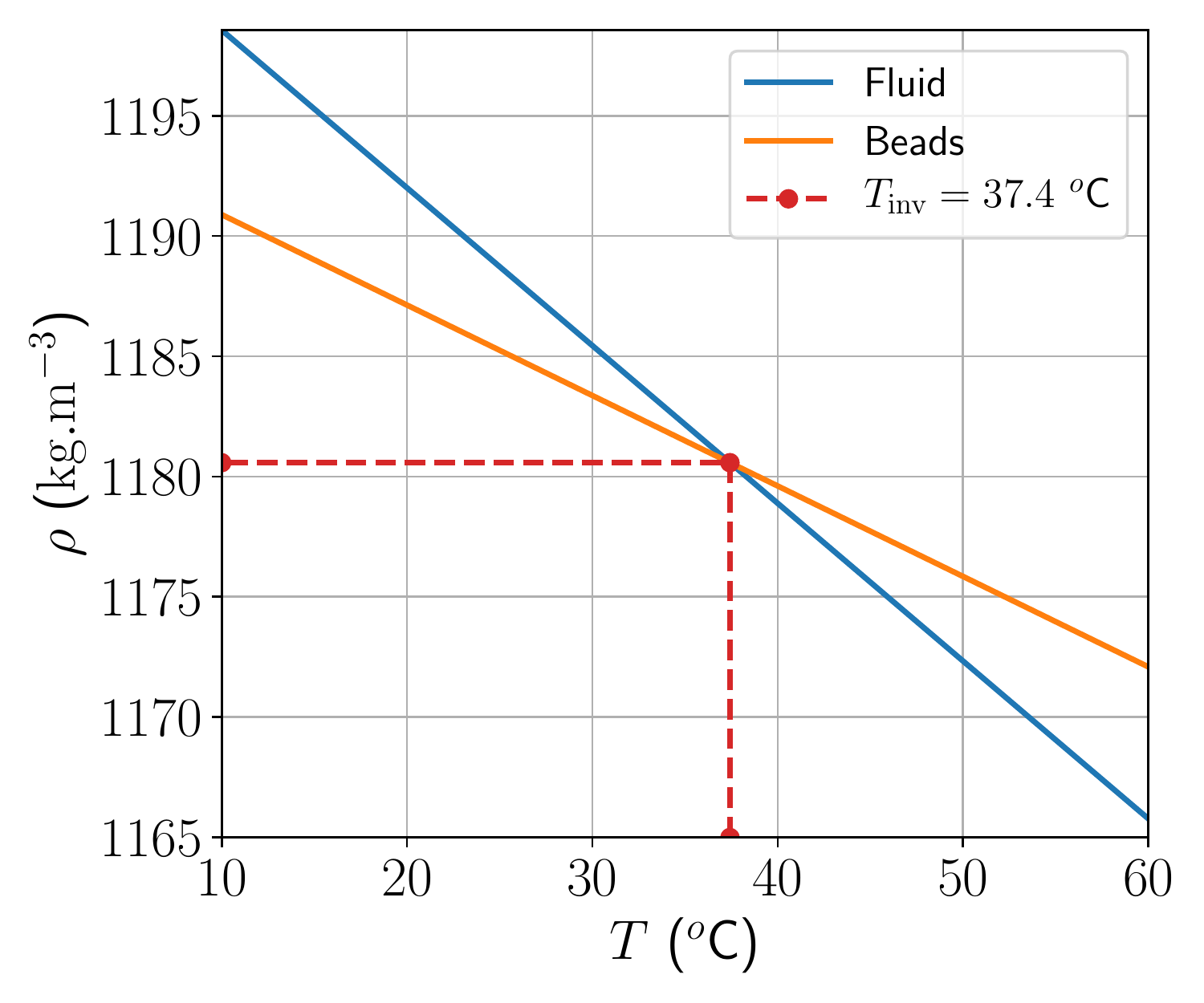}
	\caption{Variation of the density of  the beads and the fluid with temperature. The range of temperatures reached in experiments ($10-60^{o}\rm{C}$) gives rise to both sinking or floating particles behaviors.}
	\label{fig:1}
\end{figure}
	
\begin{table}
	\centering
	\begin{tabular}{c c c c}
	\hline
	Properties 					& Symbol 			&Value 		& Unit\\
	\hline
	Fluid density ($20^{o}\rm{C}$)		& $\rho_{0,f}$ 		& $1192$ 		& $\mathrm{kg\ m^{-3}}$\\   
	Beads density ($20^{o}\rm{C}$)		& $\rho_{0,p}$		& $1187$ 		& $\mathrm{kg\ m^{-3}}$\\
	Fluid thermal expansion 			& $\alpha_{f}$		& $5.5\ 10^{-4}$& $\mathrm{K^{-1}}$\\
	Beads thermal expansion 			& $\alpha_{p}$		& $3.2\ 10^{-4}$& $\mathrm{K^{-1}}$\\
	Fluid viscosity ($20^{o}\rm{C}$)		& $\eta_{f}$		& $0.151$ 	& Pa\ s\\
	Activation energy 				& $E_{a}$			& $41.7$ 		& $\mathrm{kJ\ mol^{-1}}$\\
	Fluid thermal diffusivity			& $\kappa_{f}$		& $9.1\ 10^{-8}$& $\mathrm{m^{2}\ s^{-1}}$\\
	Beads thermal diffusivity (*) & $\kappa_{p}$	& $1.\ 10^{-7}$ 	&$\mathrm{m^{2}\ s^{-1}}$\\
	Fluid thermal conductivity 			& $\lambda_{f}$	& $0.276$ 	&$\mathrm{W\ m^{-1}\ K^{-1} }$\\
	Beads thermal conductivity (*) 		& $\lambda_{p}$	& $0.21$ 		& $\mathrm{W\ m^{-1}\ K^{-1} }$\\
	\hline
	\end{tabular}
	\caption{Main physical properties of the fluid and beads. The activation energy is obtained from the viscosity fit with an Arrhenius law: $\eta(T)=\eta_{f} \exp\left[\dfrac{E_{a}}{R}\left(\dfrac{1}{T}-\dfrac{1}{T_{0}}\right)\right]$, with $T_{0}=20^{o}\rm{C}$. Properties are all measured in the lab, except those marked with (*) which are taken from \cite{HandbookPolymer}. See Supplementary Materials for further information on the way properties measurements have been carried on.}
	\label{tab:ParamsExp}
	
\end{table}

\begin{figure}
	\centering
	\includegraphics[width=0.9\textwidth]{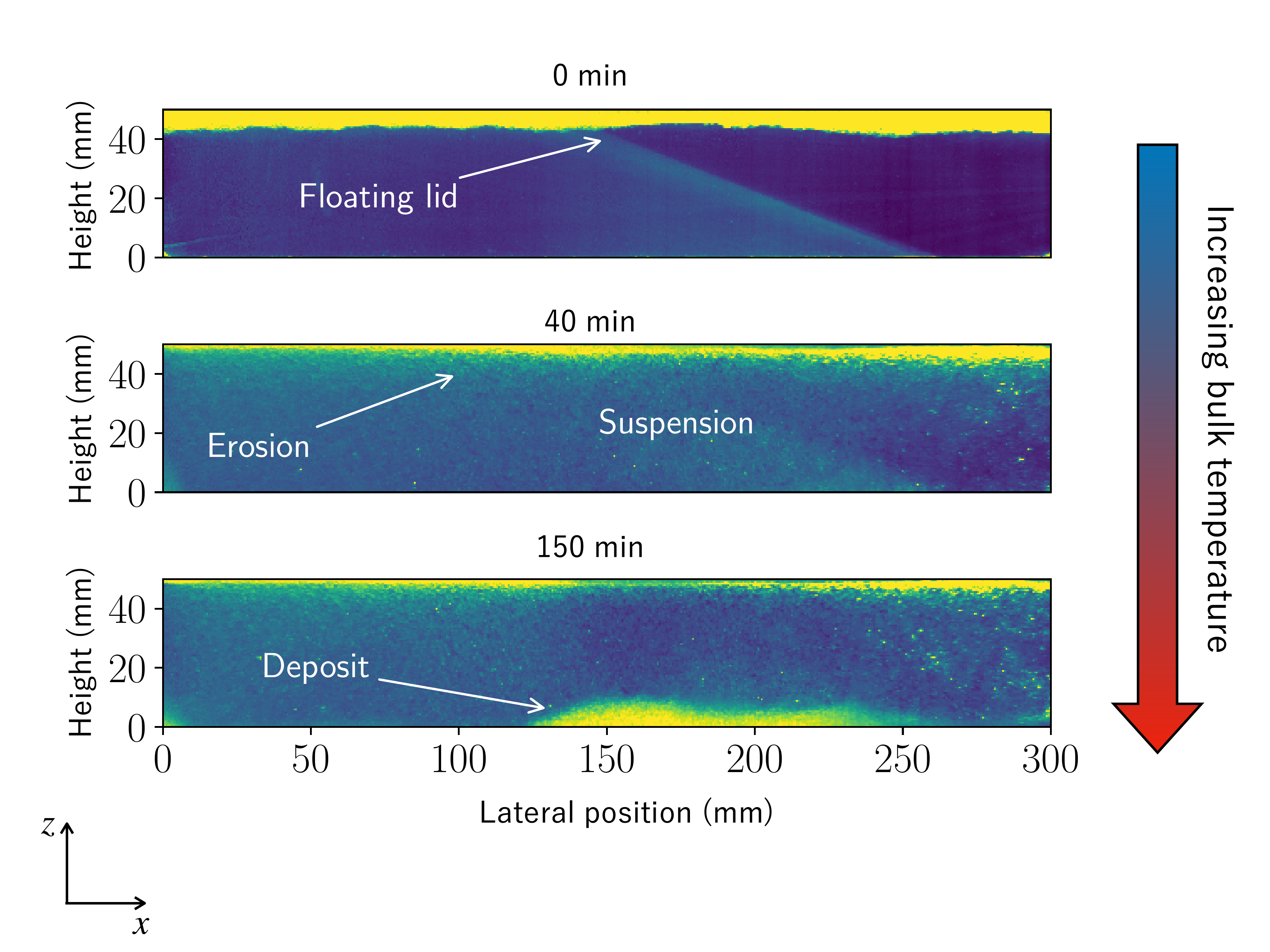}
	\caption{Snapshots of one cross-section ($y=20\, \rm{mm}$ from the tank front wall) during an experiment (IHB05), showing the typical phenomena observed. Images  are taken at 3 different moments in time: $t=0, \, 40$ and $150\, \rm{min}$ after the microwave power was turned on. Colors stand for the laser light intensity scattered. Blue color corresponds to the fluid. Hotter colors correspond to beads. At $t=0$, convection begins, and particles are eroded from the floating lid. The erosion of the floating lid can be either partial or total depending on the experimental conditions (table \ref{tab:Exp}). In some cases, due to progressive heating of fluid, beads become heavier than the fluid and settle to form a cumulate.}
	\label{fig:2}
\end{figure}


	\subsection{Fluid and particles}
	 \indent The working fluid used in experiments is a mixture of $44\ \mathrm{wt}\%$ glycerol and $56\ \mathrm{wt}\%$ ethylene glycol. Particles are PMMA spherical beads. Two sets of beads are used, corresponding to two different radii ($r_{1}=290\ \mu m$, $r_{2}=145\ \mu m$). The main properties are summarised in table \ref{tab:ParamsExp}. Particles have a different thermal expansion coefficient $\alpha_{p}$ than the fluid allowing the investigation of the full range of particle behaviours. For both phases, the density is linked to the temperature according to the thermal equation of state:
	\begin{equation}
		\rho_{i}(T)=\rho_{0,i}\left[1-\alpha_{i}\left(T-T_{0}\right)\right], \label{eq:31}
	\end{equation}
	where $T_{0}$ is the reference temperature, $\rho_{0,i}$ is the reference density at the reference temperature $T_{0}$, and $i$ refers to the fluid or particles. In this case, the density difference between the fluid and particles is:
	\begin{equation}
		\Delta \rho(T)=\rho_{f}(T)-\rho_{p}(T)=\Delta \rho_{0}\left(1-\frac{\Delta (\rho_{0}\alpha)}{\Delta \rho_{0}}\ T\right), \label{eq:32}
	\end{equation}
	 with $\Delta (\rho_0 \alpha)=\rho_{0,f}\alpha_f-\rho_{0,p}\alpha_p$. If the fluid is cold enough, particles are lighter than the fluid and float, whereas at higher temperature, beads become heavier and can sink. Thus, an inversion of buoyancy exists at an ``inversion temperature'' $T_{\mathrm{inv}}$  (figure \ref{fig:1}). Furthermore, when the thermal state in the experimental tank spans a large range of temperature, from the cold surface temperature $T_{\mathrm{s}}<T_{\mathrm{inv}}$ to the bulk temperature $T_{\mathrm{bulk}}>T_{\mathrm{inv}}$, the system can display simultaneously both a floating lid and cumulate formation. In our case, $T_{\mathrm{inv}}=37.4^{o}\rm{C}$. \\
	\indent Experimental conditions are summarized in table \ref{tab:Exp}. Fluid Prandtl number is high ($Pr\approx 1000$) and experiments reached high Rayleigh-Roberts numbers ($Ra_{H}\in[3.10^{6},\, 10^{8}]$). Particles Stokes number is about $10^{-5}-10^{-4}$, which makes them passive tracers.\\
	\indent As particles are made of PMMA, they are transparent to microwave radiation, so internal heating only occurs in the fluid. Comparing the diffusive time scale in one particle $\tau_{d}\sim r^{2}/\kappa_{p}$ and the convective time scale $\tau_{c}\sim h/W$ leads to $\tau_{c}/\tau_{d}\approx10^{1}-10^{2}$, which shows that the thermal equilibration of particles is rapid. Thus, we assess that the local temperature difference between the particles and the fluid is negligible.\\
	\indent Unfortunately, in this configuration, we could not achieve refractive index matching between the beads and particles (see Supplementary Materials). Consequently, plastic beads are light-scattering objects and the images recorded by the cameras are blurred. This has little influence on the velocity measurement, as particles behave like passive tracers. In order to check if it affects the temperature measurements, we carried out a sedimentation experiment using an homogeneous suspension in a controlled isothermal environment. We monitored the fluorescent signal whilst particles settled, and confirmed that the mean temperature measured was consistent with the imposed temperature. Hence, the presence of beads does not affect the measurement of the average properties of convection, on which further reasoning is based.

\begin{table}
   \centering
   \begin{tabular}{c  ccccccc  }
      \hline
      Name 				& Beads radius	($\mu$m)		&  $\delta_0$ (mm)	& $T_s$ ($^o$C)	& $T_{bulk}$ ($^o$C)		&	 $Ra_H$ ($10^7$)  		& Erosion?	&    Cumulate?	\\
      \hline

      IHB04         		 	& 	$290$	&		5.3			&	21.8			&	44.5		&		4.3			& Partial		& yes		\\
      IHB05		  	 	& 	$290$	&		5.3			&	21.8			&	47.7		&		8.9			& Partial		&yes			\\
      IHB07 	     	 	&	 $290$	&		1.9			&	22.3			&	35.9		&		2.8			&Partial		&no			\\
      IHB08 	     	 	&	 $290$	&		1.9			&	14			&	32.8		&		2.4			&Partial		&no			\\
      IHB09 	     	 	&	 $290$	&		1.9			&	34.1			&	43.4		&		4.1			&Partial		&yes			\\
      IHB11		    	 	&	 $290$	&		4.7			&	22.8			&	43.9		&		4.2			&Partial		&yes			\\
      IHB12		    	 	&	 $290$	&		4.7			&	22.8			&	35.3		&		1.4			&Partial		&no			\\
      IHB13 	    	 	&	 $290$	&		4.7			&	23			&	48.3		&		8.0			&Partial		&yes			\\
      IHB14		    	 	&	 $290$	&		4.7			&	8.9			&	42.0		&		5.9			&Partial		&no			\\
      IHB16				&	 $290$ 	&		3.8			&	22.3			&	53.7		&		14.4			&Partial		&yes 		\\
      IHB17				&	 $290$	&		3.8			&	22.2			&	38.7		&		2.8			&Partial		&no			\\
      IHB18				&	 $290$ 	&		3.8			&	22.2			&	47.4		&		11			&Partial		&yes			\\
      IHB18SC			&	 $290$	&		3.8			&	27			&	51.2		&		11			&Partial		&yes			\\
      IHB19				&	 $145$ 	&		4.7			&	22.9			&	42.1		&		6.8			&Total		&no			\\
      IHB20				&	 $145$ 	&		4.7			&	29.2			&	45.7		&		8.1			&Total		&no			\\
      IHB20hot			&	 $145$ 	&		4.7			&	34.5			&	49.4		&		9.7			&Total		&yes			\\
      IHB21				&	 $145$ 	&		4.7			&	23.1			&	30.6		&		1.5			&Total		&no			\\
      IHB22				&	 $145$ 	&		4.7			&	30.5			&	45.3		&		10			&Total		&yes			\\
      IHB23				&	 $145$ 	&		4.7			&	23.2			&	33.1		&		2.6			&Total		&no			\\
      IHB24				&	 $145$ 	&		4.7			&	23.1			&	26.3		&		0.27			&Partial		&no			\\
      IHB25				&	 $145$ 	&		4.7			&	23.2			&	28.5		&		0.67			&Total		&no			\\
      IHB26				&	 $145$ 	&		4.7			&	33.2			&	36.3		&		1.0			&Total		&no			\\
      IHB27				&	 $145$ 	&		4.7			&	34.4			&	43.7		&		4.4			&Total		&no			\\
      IHB27TS			&	 $145$ 	&		4.7			&	37.7			&	45.3		&		4.8			&Total		&yes			\\	
      \hline
      \hline
      IHB29\_1 			&	 -	 	&		0			&	22.2			&	26.0		&		0.28			&	-		&	-			\\	      
      IHB29\_2			&	 -	 	&		0			&	22.2			&	34.5		&		3.4			&	-		&	-			\\
      IHB29\_3 			&	 -		&		0			&	22.3			&	38.9		&		7.9			&	-		&	-			\\
      IHB30\_1    			&	 -	 	&		0			&	22.3			&	28.5		&		0.77			&	-		&	-			\\
      IHB30\_2			&	 -	 	&		0			&	22.4			&	31.4		&		1.8			&	-		&	-			\\
      IHB30\_3 			&	 -	 	&		0			&	25.6			&	43.1		&		12.2			&	-		&	-			\\
      IHB31\_1			&	-	 	&		0			&	9.1			&	16.5		&		0.16			&	-		&	-			\\
      IHB31\_2			&	-	 	&		0			&	29.4			&	46.6		&		13.3			&	-		&	-			\\
   \end{tabular}
   \caption{Experimental characteristics: the beads radius $r$, the initial floating bed thickness $\delta_{0}$, the imposed surface temperature $T_{s}$, the mean bulk temperature at steady state $T_{\mathrm{bulk}}$, the Rayleigh-Roberts number calculated at steady state $Ra_{H}$. The two last columns inform about the erosion of the floating lid (whether partial or total), and the formation of a cumulate at the bottom of the tank. Note that two families of beads with different radius are used. The last 8 rows correspond to experiments done without beads.}
   \label{tab:Exp}
\end{table}


	\subsection{Experiments}
	\indent Experiments were conducted as follows. The tank was filled with a mixture of fluid and particles. One has to avoid introducing air into the system, in order to limit surface tension effects (see Supplementary Materials for more details). The system is thermostated at the surface temperature $T_{s}$ generally bellow $T_{\mathrm{inv}}$, so that particles form initially a floating lid (see figure \ref{fig:2}). Then, the microwave power is turned on and convection starts within a time lapse of few tens of seconds. As convection proceeds, the floating lid is eroded, and particles are put in suspension. In some experiments, the inversion temperature was reached, and particles could further form a cumulate (Table \ref{tab:Exp}). We waited until the thermal steady state was reached (which corresponds to a period of time spanning from 2 to 6 hours).\\
\indent In the following section, we will study in detail these two aspects: the erosion of the floating lid and the formation of the cumulate.

	
\section{Floating lid}


	\subsection{Thermal steady state}

	\begin{figure}
	\centering
	\includegraphics[width=0.9\textwidth]{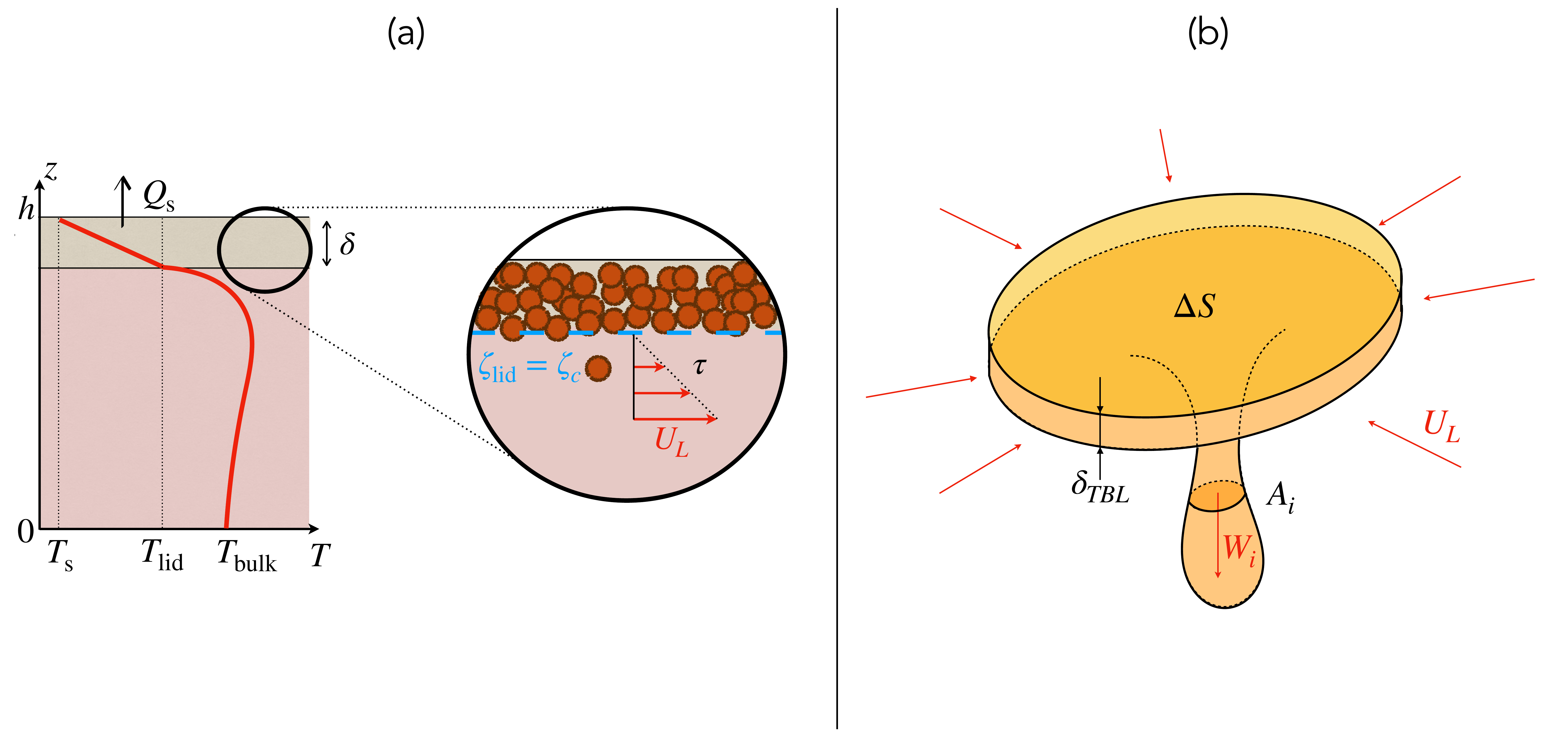}
	\caption{(a): Illustration of the thermal state of the system and of the erosion model used to determine the lid thickness at steady state, with $h$ the reservoir thickness, $T_{s}$ the surface temperature, $T_{\mathrm{lid}}$ the basal temperature of the floating lid, $T_{\mathrm{bulk}}$ the average bulk temperature, $Q_{s}$ the surface heat flux, $\delta$ the floating lid thickness, $\zeta$ the Shields number defined in the text, $\zeta_{c}$ its critical value, $\tau$ the convective shear stress, and $U_{L}$ the horizontal velocity scale. (b): Schematic view of a downwelling used for the shear stress scaling law. $\Delta S$ stands for the surface from which the fluid is drained, $\delta_{TBL}$ is the TBL thickness, $W_{i}$ the maximal velocity of cold plumes.}
	\label{fig:4}
\end{figure}
\begin{figure}
	\centering
	\includegraphics[width=0.55\textwidth]{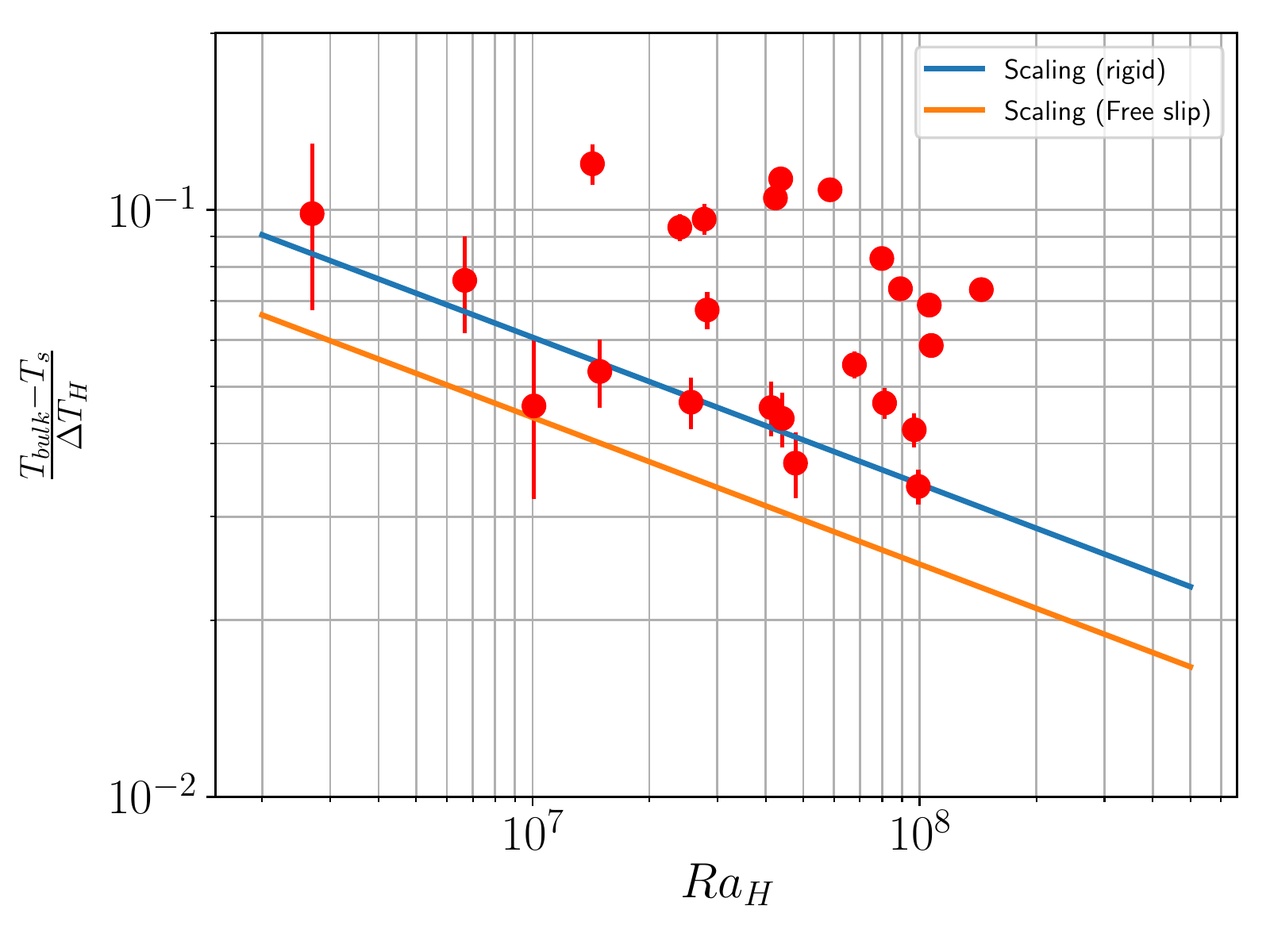}
	\caption{Experimental dimensionless drop of temperature $(T_{\mathrm{bulk}}-T_{s})/\Delta T_{H}$ as a function of the Rayleigh-Roberts number at steady state. The two solid lines stand for scaling laws for homogeneous internal heating with rigid and free-slip top conditions: $(T_{\mathrm{bulk}}-T_{s})/\Delta T_{H}=C_{T} Ra_{H}^{-1/4}$ where $C_{T}$ is given Table \ref{tab:coeff}).}
	\label{fig:3}
\end{figure}
\begin{figure}
	\centering
	\includegraphics[width=\textwidth]{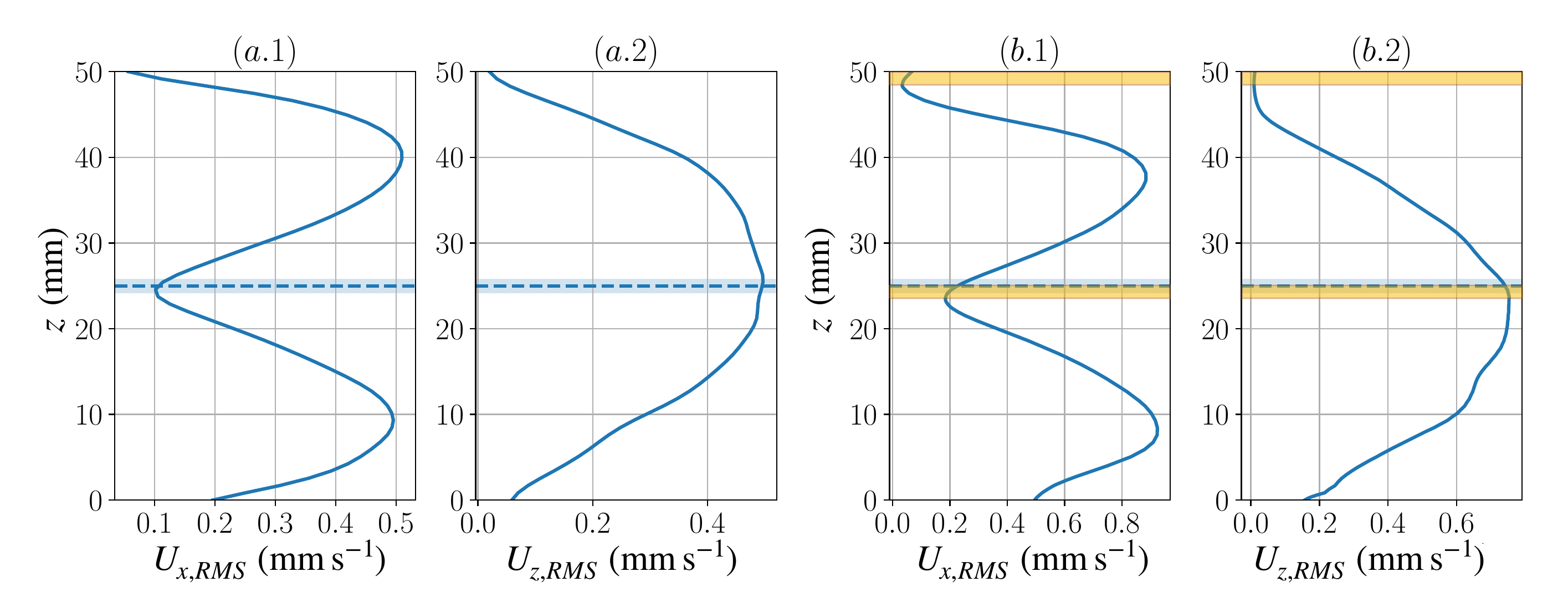}
	\caption{Root mean square horizontal and vertical velocity values at steady state, averaged at each depth $z$ for an experiment without beads $(a)$ and with a floating lid $(b)$. The mid-depth of the tank is represented by the dashed line. In the case $(b)$, the depth at which the local extrema of the velocity is shifted, because of the floating lid schematized by the yellow bands.}
	\label{fig:5}
\end{figure}
\begin{figure}
	\centering
	\includegraphics[width=0.6\textwidth]{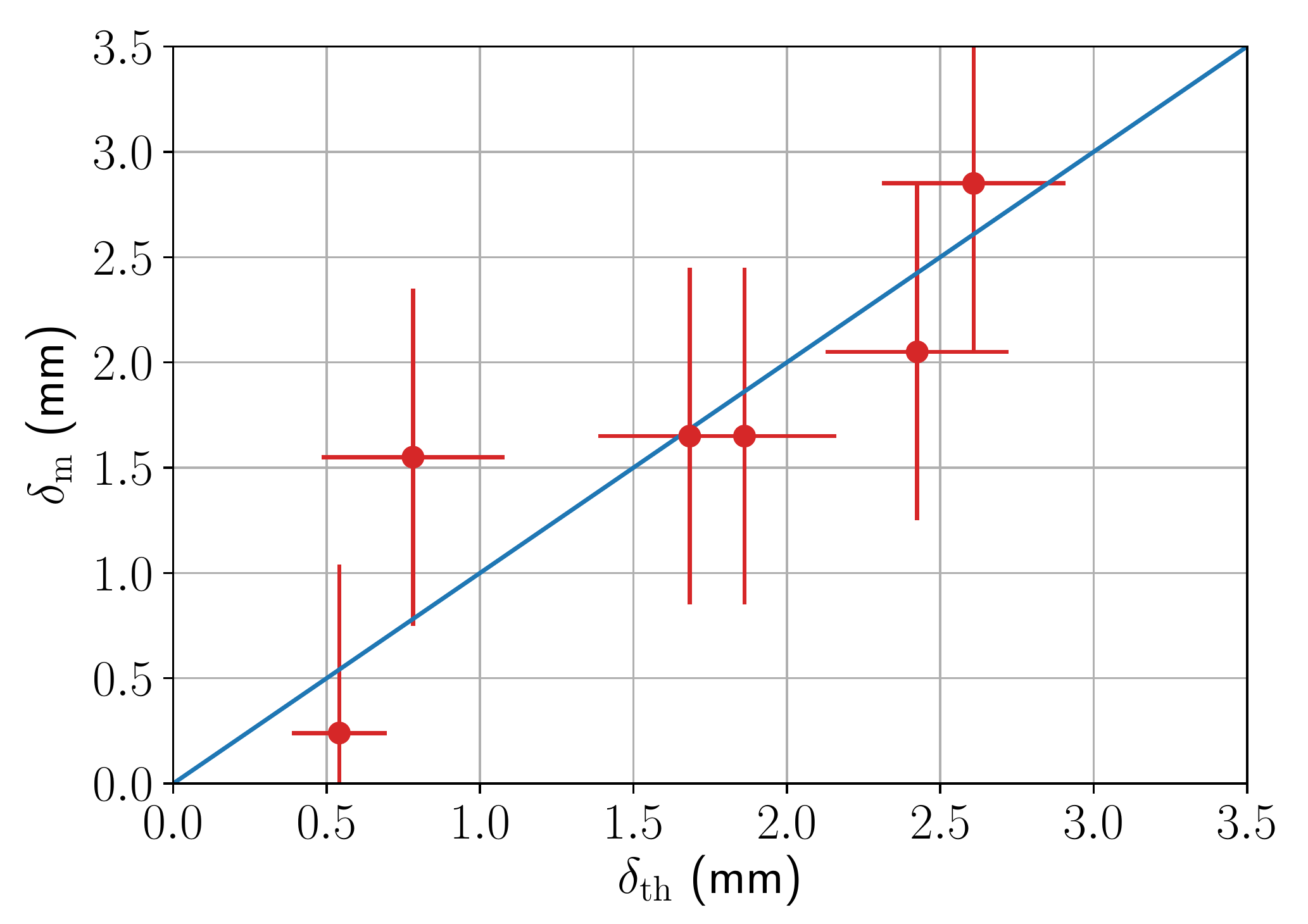}
	\caption{Comparison between the lid thickness inverted from the thermal state $\delta_{\mathrm{th}}$ and the one measured by the velocity-shift method $\delta_{m}$. Error-bars on $\delta_{m}$ correspond to one pixel of the PIV-grid, and those on $\delta_{\mathrm{th}}$ correspond to one particle radius.}
	\label{fig:6}
\end{figure}

	\indent The presence of a floating lid is likely to influence the thermal state of the system as it is situated in the TBL (figure \ref{fig:4}). In the following section, we will quantify the thickness of the steady lid and the thermal state of the system. \\
	\indent Figure \ref{fig:3} reveals that the dimensionless drop of temperature between the bulk and the surface is systematically higher than the one predicted by the scaling laws (\ref{eq:210}). We displayed the scalings for both mechanical conditions (rigid and free-slip) since this boundary condition is not well defined beneath a granular lid. The lid reduces the efficiency of heat transfer that occurs at the top of the reservoir and causes an insulating effect related to the lid thickness.\\
	\indent To quantify this effect, we use the model developed to study convection under a conductive lid \citep{Guillou95,Lenardic03,Grigne07}. The reasoning is based on two hypotheses. First, we consider the lid as a homogeneous conductive layer with an average conductivity $\overline{\lambda}$, and an averaged thickness $\delta$, such that the heat flow through the lid is: 
	\begin{equation}
		Q_{s}=\overline{\lambda}\ \frac{T_{\mathrm{lid}}-T_{s}}{\delta_{\rm{th}}}, \label{eq:41}
	\end{equation}	
	where $\mathrm{T_{\mathrm{lid}}}$ is the basal temperature of the lid and $Q_{s}=Hh$ is the surface heat flux at steady state. The lid's mean conductivity is estimated based on the individual values of each component weighted by their respective volume fraction: $\overline{\lambda}=\phi_{\mathrm{RLP}}\lambda_{p}+(1-\phi_{\mathrm{RLP}})\lambda_{f}$, with $\phi_{\mathrm{RLP}}$ the beads packing in the lid, assumed to be the random loose one: $\phi_{RLP}=56\%$. We further consider that the scaling laws for thermal convection hold true beneath the conducting lid. It has been verified experimentally for homogeneous internally heated systems that the drop of temperature between the surface and the bulk  $T_{\mathrm{bulk}}-T_{s}$ scales like the drop of temperature through the TBL $\Delta T_{TBL}$ given in (\ref{eq:210}) with a pre-factor $C_{T}=3.38\pm0.16$ at steady state \citep{Limare19}. Thus, thanks to the continuity of temperature between the lid and the fluid, one can get:
	\begin{equation}
		T_{\mathrm{bulk}}= T_{\mathrm{lid}}+C_{T}\Delta T_{H}Ra_{H}^{-1/4},   \label{eq:42}
	\end{equation}	
	which combined with (\ref{eq:41}) yields:
	\begin{equation}
		\frac{\delta_{\rm{th}}}{h}=\frac{\overline{\lambda}}{\lambda_{f}}\ \left(\frac{T_{\mathrm{bulk}}-T_{\mathrm{s}}}{\Delta T_{H}}-C_{T}Ra_{H}^{-1/4}\right), \label{eq:43}
	\end{equation} 
	 showing that the larger the lid thickness, the hotter the bulk.\\
	\indent To validate this model that uses the thermal state of the system to estimate the lid thickness, an independent way to measure $\delta$ is required. In that aim we explore the root mean square (RMS) velocity field. Figures \ref{fig:5} ($a.1$) and ($a.2$) highlight the shape of RMS velocity profiles for a convective fluid in the absence of particles. Because of rigid boundary conditions, velocities vanish at the surface and at the floor of the reservoir. Consequently, by symmetry, the vertical velocity field has a maximum value at mid-depth -- figure \ref{fig:5} ($a.2$). Also by symmetry, the horizontal profiles have a minimum value at mid-depth between two maxima corresponding to $1/4$ and $3/4$ of the total thickness of the tank as can be seen in figure \ref{fig:5} ($a.1$). This shape is due to the confined environment that creates a horizontal recirculation flow. As emphasized in figure \ref{fig:5} ($b.1$) and ($b.2$),  the presence of the floating lid shifts vertically the point where the horizontal velocity is minimal by an amount $\Delta \delta$, which is set by the ``mechanical'' thickness of the lid $\delta_{m}$:
	\begin{equation}
		\Delta \delta=\frac{\delta_{m}}{2}. \label{eq:46}
	\end{equation}
	\indent Comparison between the thickness deduced from the velocity-shift method $\delta_{m}$ and the inverted thermal thickness  $\delta_{\mathrm{th}}$ calculated thanks to (\ref{eq:43}) is shown in figure \ref{fig:6}. In this plot, we only use experiments where the lid is partially eroded and where no cumulate appears at steady state, in order to limit unwanted shifts due to the presence of a deposit of particles at the base of the tank. The agreement between the measurements and the model is fair, despite some scatter due to the simplifications of the model used. We assessed that the floating lid can be approximated by an homogeneous conductive lid whilst it is composed of packed particles containing interstitial fluid. Moreover, its thickness is  not strictly uniform as dunes form during erosion/deposition processes (see Supplementary Materials). In the following we will use the thermal thickness $\delta_{\rm{th}}$ to characterize the lid thickness.\\
	\indent Besides, the symmetry observed in the horizontally averaged vertical profile $U_{x,RMS}$ is clearly related to the return flow due to the rigid lateral boundaries and is not a general feature of convection driven by internal heating. For instance, in the case of a spherical shell, this symmetry has a priori no reason to exist. This experimental feature was used as an additional measurement of the floating lid thickness. This assumption does not affect the validity of the reasoning as our model does not require a symmetry of the lateral flow.
	 
	\subsection{Predictive model for the floating lid}

		\subsubsection{Local mechanical equilibrium}

\indent As illustrated in figure \ref{fig:4} (a), local equilibrium of the beads is set by the balance between erosion forces and the bead buoyancy. To quantify such an equilibrium, we rely on the threshold theory of mechanical erosion \citep{Glover51,Metivier17} and we define the Shields number $\zeta_{\mathrm{lid}}$ at the base of the floating lid as:
\begin{equation}
	\zeta_{\mathrm{lid}}=\frac{\tau}{\Delta \rho(T_{\mathrm{lid}})gr}, \label{eq:47}
\end{equation}
where $\tau=\eta_{f} \dot \gamma$ is the characteristic convective shear acts on the bottom of the lid, and $\dot \gamma$ the corresponding strain rate. We consider only the experiments with partial erosion of the floating lid at steady state, which means that the lid thickness $\delta \neq 0$. In these experiments, the Shields number reaches the critical threshold ($\zeta_{\mathrm{lid}}=\zeta_{c}$ in (\ref{eq:47}) and $\delta=\delta_{\mathrm{th}}$ in (\ref{eq:41})). At steady state, we assume that the temperature filed in the floating lid varies linearly with depth, so that the temperature at the base of the lid is $T_{\rm{lid}}=T_s+Q_s \delta_{\rm{th}} /\overline \lambda$. Introducing this temperature in (\ref{eq:32}), some algebra yields the following set of equations:
\begin{eqnarray}
	\delta_{c}&=&\delta^{*}\left(1-\frac{\zeta_{s}}{\zeta_{c}}\right),       \label{eq:48}\\
	\delta^{*}&=&\frac{\Delta \rho (T_{s})}{\Delta (\rho_{0}\alpha)}\ \frac{\overline{\lambda}}{Q_{s}},   \label{eq:49} \\
	\zeta_{s}&=&\frac{\tau}{\Delta \rho(T_{s})gr},      \label{eq:410}
\end{eqnarray}
where $\zeta_{s}$ is the Shields number calculated at the surface temperature $T_{s}$. With (\ref{eq:48}), $\zeta_s$ appears as the control parameter that determines the critical thickness $\delta_{c}$, as $T_{s}$ and $Q_{s}$ are known. The problem thus boils down to the determination of the characteristic convective shear stress $\tau$. For instance, in experiments done by \cite{Charru04}, the shear is experimentally controlled and homogeneously applied to the bed, which facilitates its characterization. Here, the bed lies in the unstable cold TBL. Thus  the flow field is complex and contains spatial and temporal fluctuations. Hence, characterizing the shear stress in a homogeneously heated convective system is required.


		\subsubsection{Scaling laws for velocities and shear stresses}

\begin{figure}
	\centering
	\includegraphics[width=0.9\textwidth]{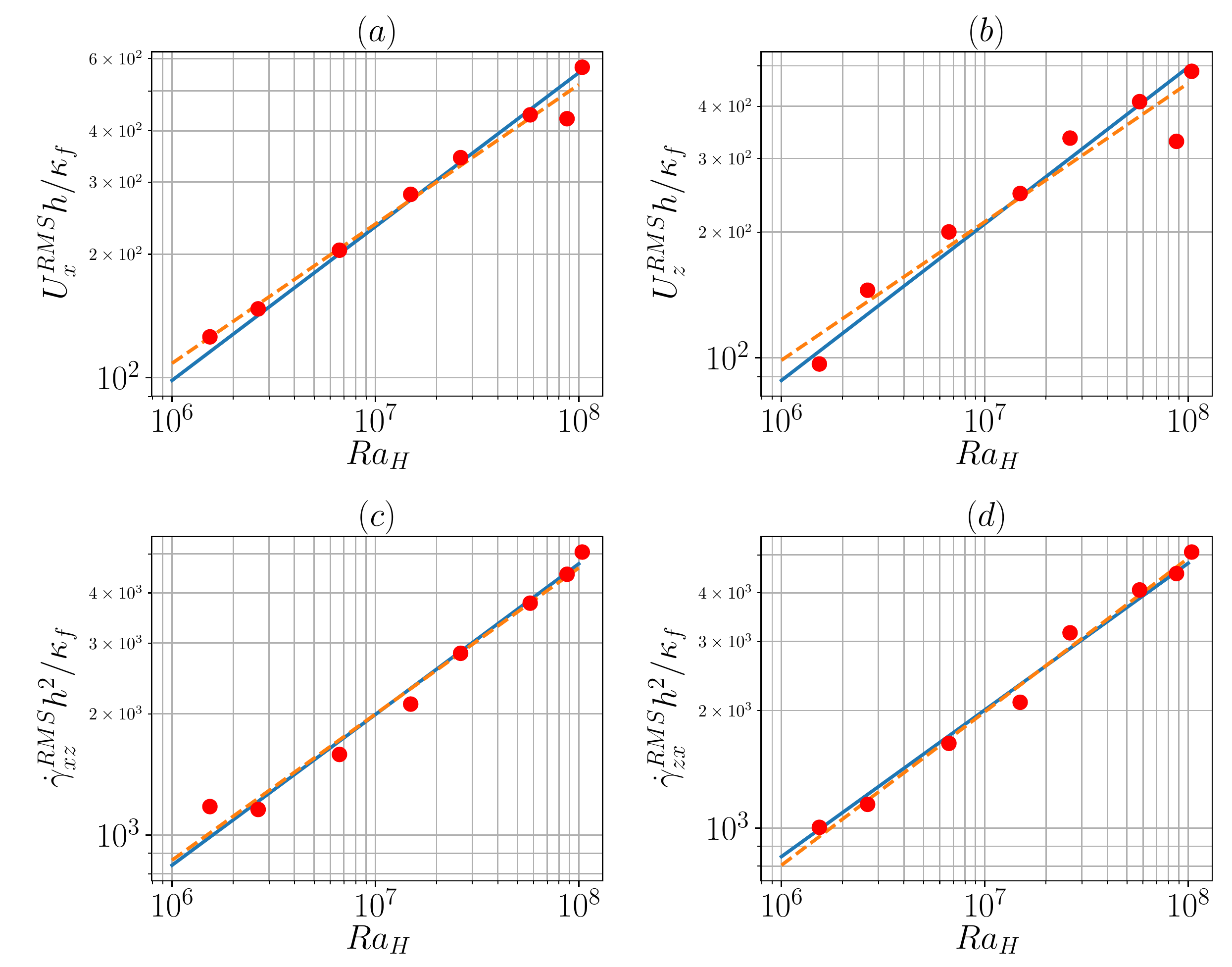}
	\caption{Scaling laws derived from experiments without beads. Horizontal $(a)$ and vertical $(b)$  velocity, horizontal $(c)$ and vertical $(d)$ strain rate. All these physical properties are evaluated thanks to their root mean square value calculated in the entire volume of the tank and are represented by the dots. Blue lines represents best fit law with fixed exponent $3/8$ and the orange dashed line are those with the exponent left to vary. All parameters of these laws are summarised in Table \ref{tab:scalings}.}
	\label{fig:8}
\end{figure}	

\indent Equation (\ref{eq:47}) emphasizes the importance of the horizontal shear stress, hence of the velocity field, on the erosion process. The strain rate $\dot \gamma$ scales as follows:
\begin{equation}
	\dot \gamma \sim \frac{U_{L}}{\delta_{v}},  \label{eq:411}
\end{equation}
where $U_{L}$ is the characteristic horizontal velocity, and $\delta_{v}$ is the characteristic length over which velocity vanishes. By definition, the latter corresponds to the dynamical boundary layer (DBL) thickness. First, the Reynolds numbers $Re$ reached in our experiments are low which implies laminar flows and thus, $\delta_{v}\sim h$ (see appendix \ref{App1}). As a consequence, the strain rate $\dot \gamma$ becomes:
\begin{equation}
	\dot \gamma \sim \frac{U_{L}}{h}. \label{eq:415}
\end{equation}
We consider the volume of fluid in the TBL that is drained by one downwelling (Figure \ref{fig:4} $(b)$). On one side, fluid from the TBL is drained at the characteristic velocity $W_{i}$ by the downwelling whose cross section area  is $A_{i}$. On the other hand, fluid is converging at the characteristic horizontal velocity $U_{L}$ through the lateral surface of the cylinder of thickness $\delta_{TBL}$ and area $\Delta S$. This reasoning is based on the fact that the fluid drained in one downwelling comes mainly from the thermal boundary layer. This assumption is valid on high-Prandtl limit where entrainment between downwellings and the ambiant fluid is negligible \citep{Davaille05}. Mass conservation imposes:
		\begin{equation}
			U_{L}\delta_{TBL}\Delta S^{1/2}\sim W_{i}A_{i}. \label{eq:412}
		\end{equation}		
		 Using the same scalings as in \cite{Vilella18}: $\delta_{TBL}\sim hRa_{H}^{-1/4}$, $\Delta S \sim h^{2} Ra_{H}^{-1/4}$, $A_{i}\sim h^{2}Ra_{H}^{-3/8}$, $W_{i}\sim \kappa_{f}/h\ Ra_{H}^{3/8}$, valid for $10^{6} \leq Ra_{H}\leq 10^{9}$, we obtain the horizontal velocity scale:
		\begin{equation}
			U_{L}=C_{u} \frac{\kappa_{f}}{h}Ra_{H}^{3/8},         \label{eq:413}
		\end{equation}
		with $C_{u}$ a pre-factor which  depends on the boundary conditions.\\ 		
\indent To verify these scaling laws, experiments without beads have been carried out using the same fluid and the same methods as those described previously. We recorded the horizontal and vertical velocities using the PIV method, and we calculated the horizontal and vertical strain rate. As we are interested in the average shear rate, we determined their RMS values calculated over the entire volume of the tank. Results are displayed in figure \ref{fig:8} $(a)$, $(b)$ for the horizontal and vertical velocities. The scaling laws pre-factors determined experimentally are summarized in table \ref{tab:scalings}. In our experiments, $Re\approx 10^{-1}-1$, so the convection is laminar. \\
The scaling law for the strain rate is therefore $C_{\gamma}\frac{\kappa_{f}}{h^{2}}\, Ra_{H}^{3/8}$. Results are displayed in figure \ref{fig:8} $(c)$ and $(d)$ for the horizontal and vertical strain rates respectively. In both cases, the predicted scaling law is in good agreement with experimental data. \\
\indent We can thus estimate the Shields' number as follows:
		\begin{equation}
			\zeta_{s}=\frac{\eta_{f} \kappa_{f}}{h^{2}\Delta \rho(T_{s}) g r}\ Ra_{H}^{3/8}. \label{eq:416}
		\end{equation}

\begin{table}
	\centering
	\begin{tabular}{c c c}
		\hline
		Variable 								& Exponent left to vary		& Fixed exponent\\
		\hline
		$U_{x}^{RMS}h/\kappa_{f}$				& $0.99\ Ra_{H}^{0.34}$		& $0.55\ Ra_{H}^{3/8}$\\
		$U_{z}^{RMS}h/\kappa_{f}$				& $1.01\ Ra_{H}^{0.33}$		& $0.50\ Ra_{H}^{3/8}$\\
		\hline
		$\dot \gamma_{xz}^{RMS}h^{2}/\kappa_{f}$	& $5.69\ Ra_{H}^{0.36}$		& $4.73\ Ra_{H}^{3/8}$\\
		$\dot \gamma_{zx}^{RMS}h^{2}/\kappa_{f}$	& $3.56\ Ra_{H}^{0.39}$		& $4.76\ Ra_{H}^{3/8}$\\
		\hline
	\end{tabular}
	\caption{Parameters of power laws determined experimentally for the horizontal and vertical RMS velocities and the horizontal and vertical RMS shear rates.}
	\label{tab:scalings}
\end{table}


		\subsubsection{Critical Shields number and stability of deposits}

\begin{figure}
	\centering
	\includegraphics[width=0.5\textwidth]{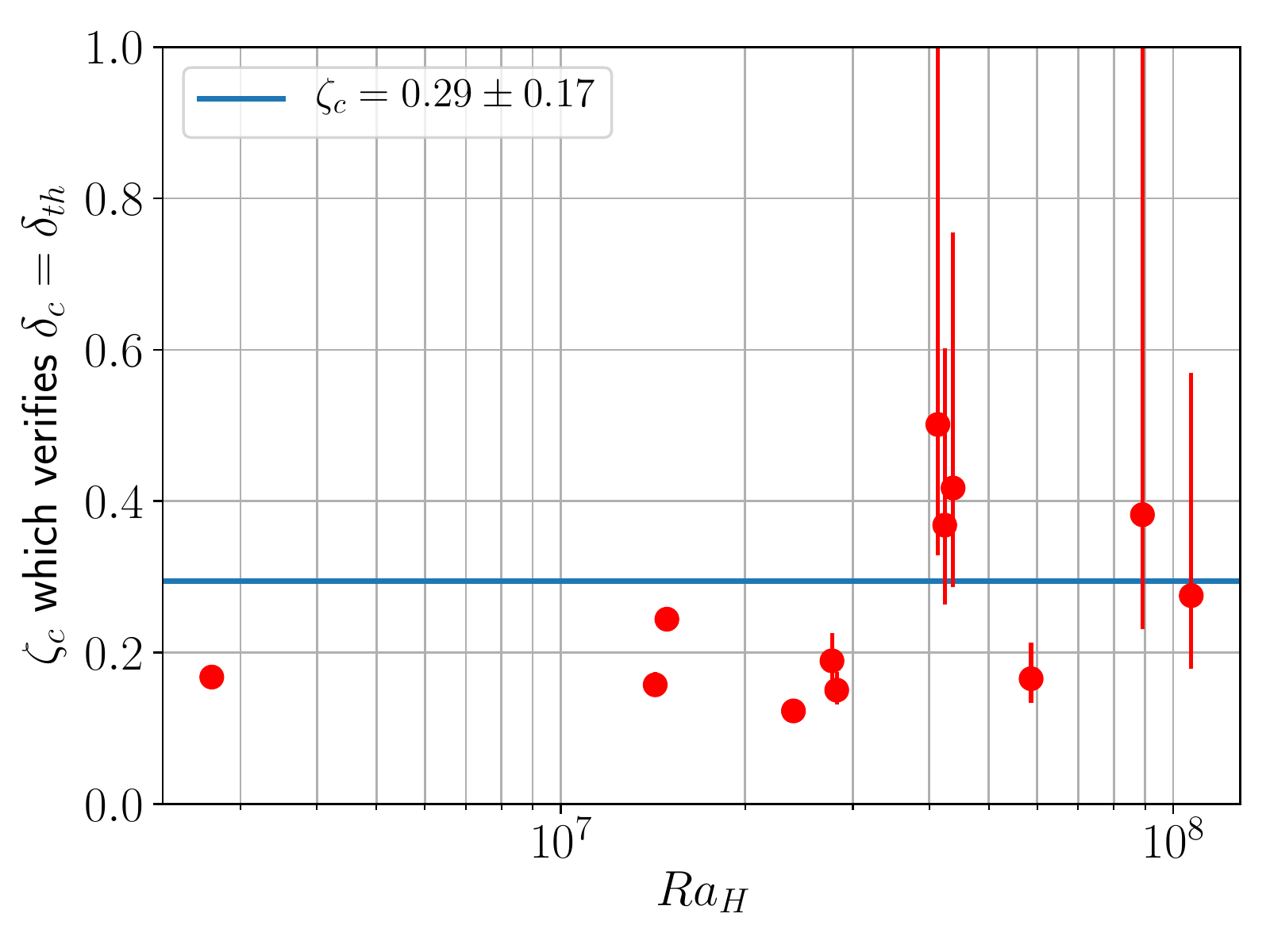}
	\caption{Determination of the critical Shields number: $\zeta_{c}=0.28\pm0.05$ to achieve a perfect match between the reference lid-thickness $\delta_{\mathrm{th}}$ and the critical thickness $\delta_{c}$.}
	\label{fig:9}
\end{figure}	

\begin{figure}
	\centering
	\includegraphics[width=\textwidth]{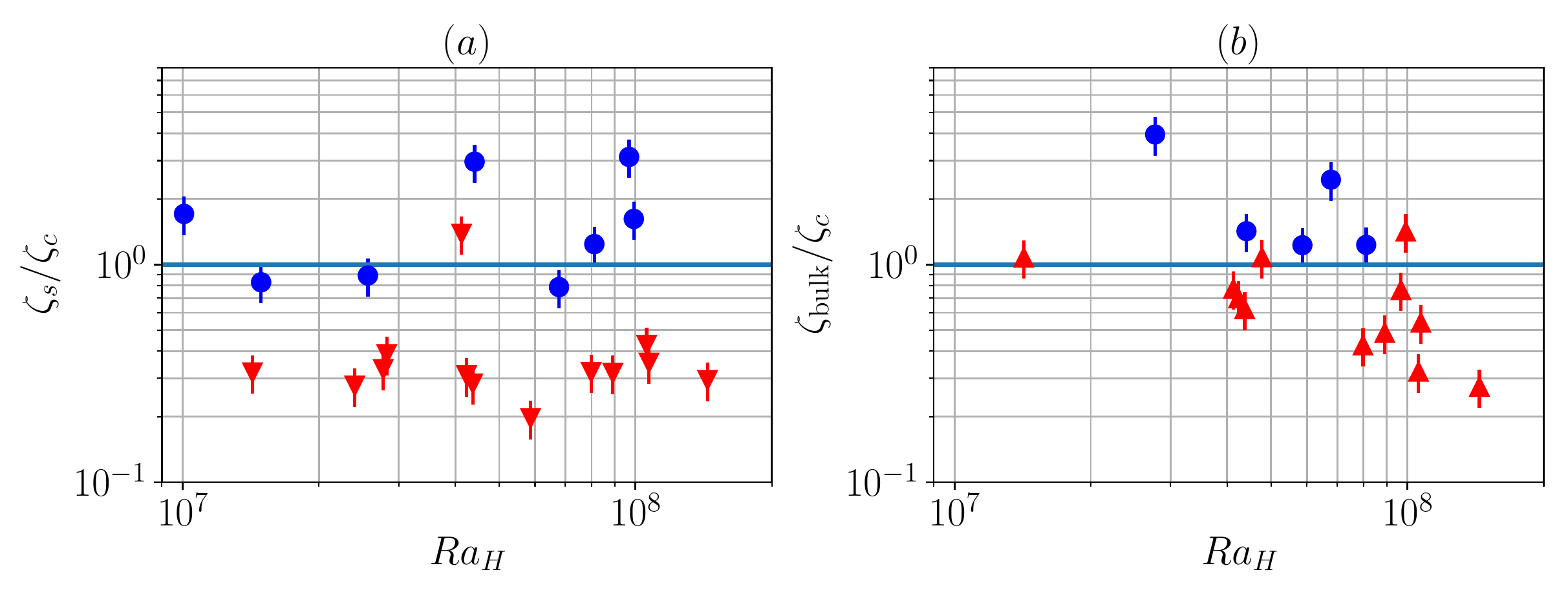}
	\caption{Stability of different deposits. $(a)$ Stability diagram for the floating lid, based on the value of $\zeta_{s}=\zeta(T_{s})$. Circles: the lid is totally eroded. Downward-pointing triangles: a stable lid is observed at steady state. $(b)$: Stability of the cumulates that is likely to form when $T_{\mathrm{bulk}}>T_{\mathrm{inv}}$. This stability is compared to the value of the local Shields number at the bottom of the tank $\zeta_{\mathrm{bulk}}=\zeta(T_{\mathrm{bulk}})$. Circles: no deposit at steady state. Upward-pointing triangles: a cumulate is observed at steady state.}
	\label{fig:10}
\end{figure}

		\indent Thanks to the previous scaling analyses, equation (\ref{eq:48}) provides a way to measure experimentally the critical Shields number. By considering that the lid thickness at steady state corresponds to the critical thickness $\delta_{\mathrm{th}}=\delta_{c}$, one can determine the critical Shields number $\zeta_{c}$ for each experiment:
\begin{equation}
	\zeta_{c}=\zeta_{s}\left(1-\frac{\delta_{\mathrm{th}}}{\delta^{*}}\right)^{-1}, \label{eq:417}
\end{equation}
with $\zeta_{s}$ given by (\ref{eq:416}). The critical number is calculated for each experiment, and results are displayed in figure \ref{fig:9}. We get: $\zeta_{c}=0.29 \pm 0.17$. This value is of the same order of magnitude of the one estimated by \cite{Charru04}. Error bars represent a variation of one bead radius of the floating lid thickness. The discrepancy is due to the sensitivity of the prediction to the value of the lid thickness, and the higher the heat flux, the steeper the thermal gradient in the lid and the greater the error bars. This is the reason why one experiment in particular (IHB13) does not appear in the plot because the uncertainties are too large.\\
		\indent  With the critical number $\zeta_{c}$ determined, we can compare the stability of the different deposits predicted by the Shields approach with the experimental observations (figure \ref{fig:10} $(a)$ and $(b)$). For the floating lid, we calculate the surface Shields number $\zeta_{s}$. If $\zeta_{s}>\zeta_{c}$, the floating lid is unstable and erosion should be total. But if $\zeta_{s}<\zeta_{c}$, a floating lid of some thickness is stable. Results are displayed in figure \ref{fig:10} $(a)$ and the transition between total erosion and partial erosion is well described by $\zeta_{c}$. Similarly, we calculate the bulk Shields number $\zeta_{\mathrm{bulk}}$, which is also the Shields number at the base of the reservoir:
		\begin{equation}
			\zeta_{\mathrm{bulk}}=\frac{\eta_{f} \kappa_{f}}{h^{2}\Delta \rho(T_{\mathrm{bulk}}) g r}\ Ra_{H}^{3/8}, \label{eq:418}
		\end{equation}
and we compare it to $\zeta_{c}$. If $\zeta_{\mathrm{bulk}}>\zeta_{c}$, convection prevents deposition at the base of the tank. Otherwise, a basal deposit is stable. This transition is also verified experimentally in figure \ref{fig:10} $(b)$.

\section{Suspension stability}\label{sec:NRJ_balance}
\begin{figure}
	\centering
	\includegraphics[width=0.5\textwidth]{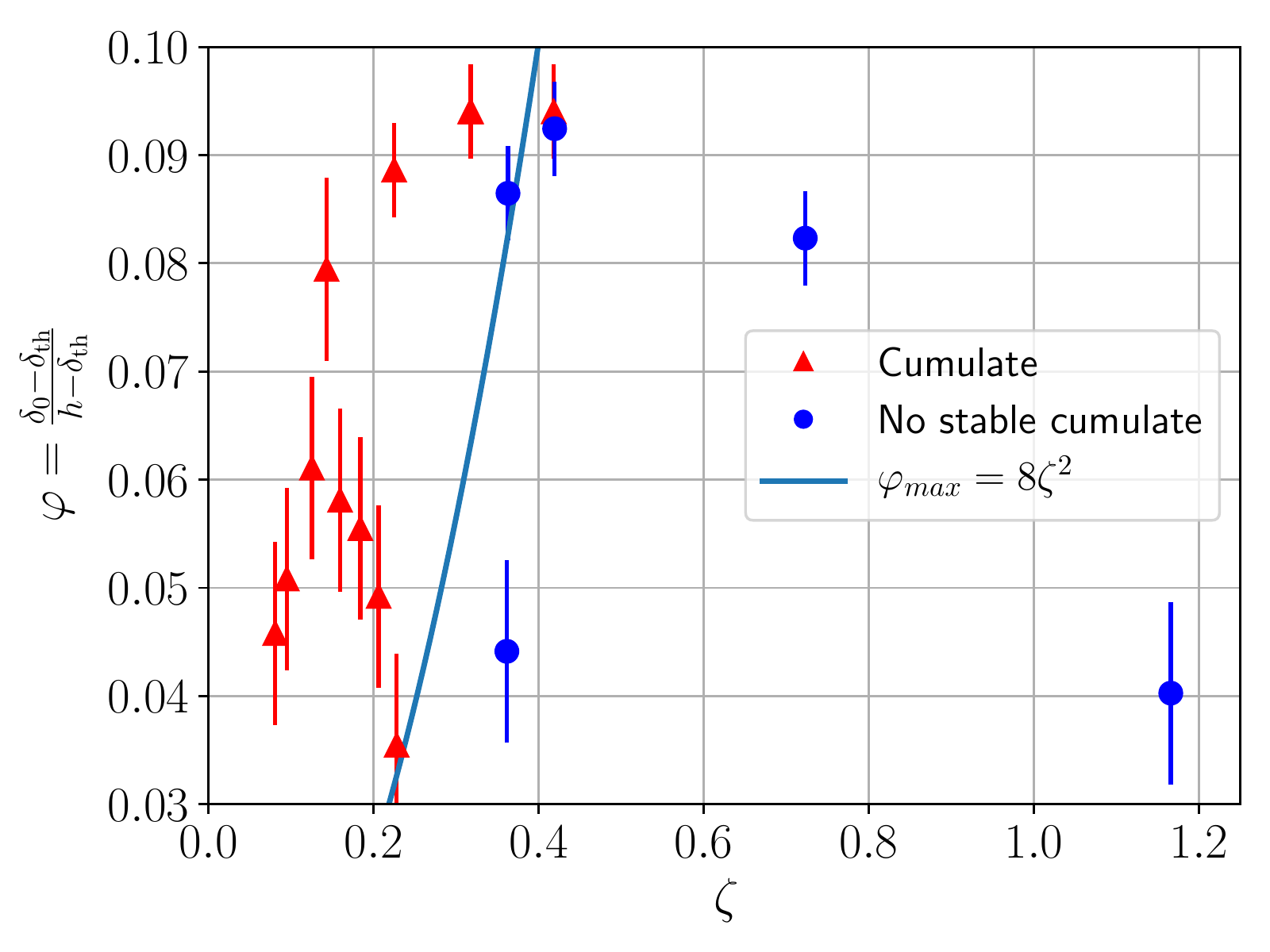}
	\caption{Proxy of particle concentration at steady state for all experiments where particles in suspension are heavier than the fluid. The blue line represent the transition between the cumulate regime et the absolute suspension one, assuming a constant efficiency coefficient.}
	\label{fig:11}
\end{figure}

	\indent The scalings derived above allow predicting whether a deposit can form at the base of the reservoir and a stable lid at the top of it, but we need another framework to describe the full dynamics of the suspension containing the particles eroded from the deposits. In a fluid at rest, particles with negative buoyancy will all eventually settle down. Observations show that in a convective fluid, even negatively buoyant particles can remain in suspension at steady state \citep{Lavorel09}. \cite{Solomatov93b} proposed that the dynamics of the suspension can be described based on the equilibrium between buoyancy and shear forces, described by the balance:
	\begin{equation}
		\phi_{max}\ \iiint_{(V)}\Delta \rho \mathbf{g}\cdot\mathbf{u}_{p}\mathrm{d}V \sim \epsilon\ \iiint_{(V)}\eta_{f} \mathbf{u}_{f}\cdot \nabla^{2} \mathbf{u}_{f}\mathrm{d}V,                \label{eq:419}
	\end{equation}
	or in compact form,
	\begin{equation}
		\phi_{max}\mathcal{B}_{p} \sim \epsilon \mathcal{V}_{f}, \label{eq:420}
	\end{equation}
	where $\mathcal{B}_{p}$ is the integral on the left-hand side of (\ref{eq:419}) related to beads buoyancy, $\mathcal{V}_{f}$ is the integral on the right-hand side of (\ref{eq:419}) referring to the bulk viscous dissipation, $(V)$ is the total volume of the reservoir, and $\epsilon$ is the percentage of viscous energy being used to maintain particles in suspension (also called \textit{efficiency parameter}). This description can be further used to determine the maximal concentration of particles $\phi_{\mathrm{max}}$ that can be maintained in suspension by convection. Assuming $\epsilon$ to be constant, \cite{Solomatov93b} get the following law for $\phi_{max}$:
	\begin{equation}
		\phi_{\mathrm{max}}= C_{s}\epsilon \left(\frac{\overline{\tau}}{\overline{\Delta \rho} g r}\right)^{2}=C_{s} \epsilon \overline \zeta^{2}, \label{eq:421}
	\end{equation}
	where $\overline{\tau}$ and $\overline{\Delta \rho}$ stand for the volume averaged values of $\tau$ and $\Delta \rho$ respectively, and $C_{s}=9/2$ \citep{Solomatov93b}. Basically, if the concentration of particles in the convective bulk $\overline \phi$ is below this limit, particles stay in suspension. Otherwise, the convective fluid only sustains the quantity of particles corresponding to the maximal concentration $\phi_{max}$, and the rest settles, forming a cumulate. Interestingly, this law can be expressed with a Shields parameter $\overline \zeta$ similar to the one used previously. The major difference lies in the fact that $\overline \zeta$ is a global parameter, comparing volume average properties, whereas previously $\zeta$ has been estimated locally. However, by neglecting the thin TBL at the top of the reservoir, one can approximate that $\overline \zeta \approx \zeta_{\mathrm{bulk}}=\zeta(T_{\rm{bulk}})$, and thus $\phi_{\mathrm{max}}$ is described by previous scaling laws. \\
	\indent In order to verify the validity of this criterion in our experiments, we considered the formation of the cumulate at the base of the tank. As the direct measurement of particle concentration $\phi$ is not possible due to the refractive index mismatch, we calculate a proxy $\varphi$ of the quantity of particles that is eroded from the floating lid. To do so, taking $\delta_{0}$ as the initial thickness of the bed, and $\delta_{\mathrm{th}}$ as the thickness at steady state, the quantity of particles that is eroded is proportional to $\delta_{0}-\delta_{\mathrm{th}}$. The coefficient of proportionality is linked to the packing of particles inside the floating lid. The packing is assumed to be constant, as all experiments are prepared similarly. Thus, the concentration of particles in the bulk is calculated as if all the eroded material stay in suspension:
\begin{equation}
	\overline \phi=a\varphi =a\frac{\delta_{0}-\delta_{\mathrm{th}}}{h-\delta_{\mathrm{th}}}, \label{eq:422}
\end{equation}
	where $a$ is a constant which depends potentially on the packing of beads inside the lid ($a=0.56$ for a random loose packing). As the measurement of the total quantity of deposit particles is subject to large uncertainties, we detect $\phi_{\rm{max}}$ as the limit between partial sedimentation and absolute suspension defined by the limit $\phi_{\mathrm{max}}\sim \overline{\zeta}^{2}$. The blue line in figure \ref{fig:11} represents the empirical boundary between the two regimes. Now, we have a complete framework that describes the steady state of any suspension in a convective environment driven by internal heating. The Shields's formalism enables us to quantify the thickness of the lid located in a boundary layer, and Solomatov's approach enables us to describe the maximal quantity of particles that can stay in suspension.

\label{sec:solomatov}

\section{Lids formation in a convective system bearing particles}

\begin{figure}
	\centering
	\includegraphics[width=0.95\textwidth]{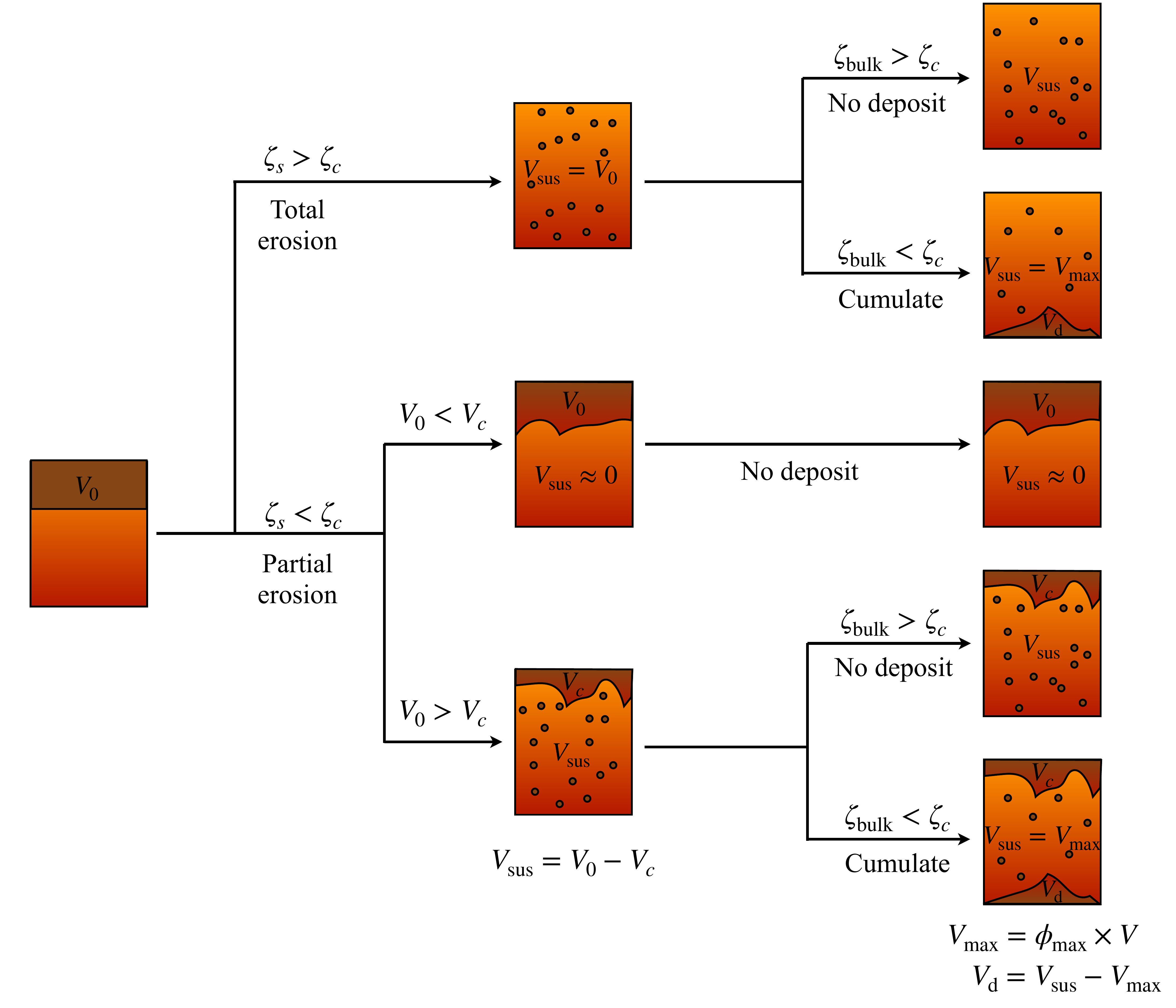}
	\caption{Synopsis that  fully determines  the system's regime at steady state. The surface Shields number $\zeta_{s}$ is given by (\ref{eq:416}) and depends on the surface temperature $T_{s}$, the Rayleigh-Roberts number $Ra_{H}$ and the initial volume of particles $V_{0}$ which are all input parameters. $V_{c}$ is the volume of the critical floating lid that can form at the surface of the convective fluid, corresponding to a lid of thickness $\delta_{c}$ given by (\ref{eq:48}). \textbf{Stability of the floating lid:} if $\zeta_{s}>\zeta_{c}$, the floating lid is unstable and all particles are put in suspension. If $\zeta_{s}<\zeta_{c}$, the erosion is partial. In this case, if $V_{0}<V_{c}$, the initial crust is stable, and very few particles are put in suspension. If $V_{0}>V_{c}$, only the volume of particles $V_{c}$ stay at the surface, the rest of the volume $V_{\mathrm{sus}}=V_{0}-V_{c}$ is put in suspension. \textbf{Cumulate formation :} we consider the stability of the suspension in the case of negatively buoyant suspended particles. If the basal Shields number $\zeta_{\rm{bulk}}$ is greater than $\zeta_{c}$, the suspension is stable and no cumulate forms. If $\zeta_{\rm{bulk}}>\zeta_{c}$, a basal cumulate can settle down. The volume of this cumulate is given by the maximal concentration of particle that can stay in suspension $\phi_{\rm{max}}$ given by (\ref{eq:421}). The convective fluid can only sustain a maximal volume $V_{\mathrm{max}}=\phi_{\rm{max}}\, V$ in suspension, where $V$ is the volume of the convective layer. $V_{\rm{max}}$ has to be compared to the volume of particle that have been eroded $V_{\rm{sus}}$. The surplus $V_{d}=V_{\mathrm{sus}}-V_{\mathrm{max}}$ forms a basal cumulate.}
	\label{fig:12}
\end{figure}

\indent Our results can be used to describe the fate of particles in a convective fluid by splitting the system into three reservoirs: (i) the floating lid situated in the TBL, (ii) the bulk fluid containing suspended particles, and (iii) the deposit at the base. Our model quantifies the volume of particles that can be stored in each reservoir (figure \ref{fig:12}). Buoyant crystals first fill the floating lid reservoir. We can define the maximal capacity of the floating lid $V_{c}$ determined by (\ref{eq:48}). Particles exceeding this critical volume of the lid remain in suspension or form a cumulate.  In the same way, for negatively buoyant crystals, if the bulk Shields number is below the critical value, a deposit may form. The concentration of crystals that stay in suspension is limited to $\phi_{\mathrm{max}}$, the rest settles and forms the basal deposit.

\section{Discussion}


\subsection{Validity of the experimental results}

\indent The approach adopted here takes into account the main mechanisms that deal with suspension and deposits stability. The main goal was to highlight how to use the framework treating of fluid/crystal interaction in the case of convective systems bearing crystals. It is based on hypotheses that allows to express the problem in terms of scaling laws that can easily be re-scaled to conveniently describe a geophysical system as it will be illustrated in next section. \\
\indent Even though experimental results are well described by our models at first order, certain discrepancies exist and should not to be ignored. They result from different factors. First, our experiments enable the estimation of the floating lid thickness by an indirect measurement based on the average temperature state of the system. This method might increase the uncertainty on the determination of $\delta$ and, thus, all parameters that are linked to it, such as the entrainment threshold. One way to improve this measurements could be the use of fluid and beads that have a near-perfect optical index matching. In this way, the bottom of the lid would be directly observable, and the thickness measurable more precisely. This prerequisite enabled the study of the bed surface motions and even of the motions of beads inside a granular bed \citep{Mouilleron09,Houssais15}. In this way, the bottom of the lid will be directly observable, and the thickness measurable more precisely. \\
\indent Second, our theory is based on averaged physical properties, which leads to the general predicament of describing a 3D problem by a 1D theory. This crude approximation leads to results that are fairly consistent with data at first order, but a local description of the convection and the erosion mechanism can improve this approach. Downwellings develop from the TBL, which can have an influence on the local behavior of beads at the bottom of the floating lid. Equivalently, we describe the floating lid as a quiescent solid medium of homogeneous thickness, but this assumptions is a strong simplification. We point out the existence of a topography (see supplementary materials) that can have an influence on the local flow in the TBL, which, in return, impacts the erosion mechanism that occurs at the interface \citep{Charru06}. Local recirculation flow that induces dune formation are intimately linked to downwellings in our case, and understanding the interaction between dune growth and downwelling dynamics can upgrade the present study. Furthermore, the suspension is also considered in our model as evenly distributed in the bulk. This assumption represent only a particular case and numerical simulations underline that crystals can concentrate depending on the flow field and their buoyancy \citep{Patocka20}. Further investigations are required to observe experimentally the behavior of particles in suspension as a function, for instance, of the Shields number, with a set-up that allow the study of a large range of values for $\zeta$.\\
\indent Dealing with the erosion mechanism, we only treated the threshold in terms of one critical value of the Shields' number. Our estimate is consistent with data but the deviation from this value can be due to an oversimplifications. First, as timescales to reach the steady state are long, the steady state might not be reached for the erosion of the floating lid when we stopped the experiment. This might induce an over-estimate of the equilibrium thickness of the bed, and consequently an under-estimate of the threshold value. Besides, the Shields number may depend on parameters such as the grain size and the flow regime. \cite{Buffington99} showed for instance a variation for turbulent flow between $0.03-0.1$, whereas \cite{Ouriemi07} underlined that the threshold is stable in laminar flow but it depends on the way the bed is packed \citep{Agudo12}. The latter authors measured changes in the critical Shields number of a factor 2 up to a factor 5 depending on the exposure of the beads to the flow and the packing and orientation of the substrate. Thus, a dependance of $\zeta_c$ on other physical parameters related to the flow (such as the $Ra_H$), or related to the substrate (such as the packing) has to be verified with an experimental device that enables a study over several orders of magnitude. \\
\indent All these effects could explain the outliers that appears in our data. Nonetheless, our results are proposed here as a first step in the path of using the physical framework used in granular media in order to describe the behavior of particles in convective systems.

	
	\subsection{Characterization of fluid/particle coupling in our experiments}
	
\begin{figure}
	\centering
	\includegraphics[width=0.5\textwidth]{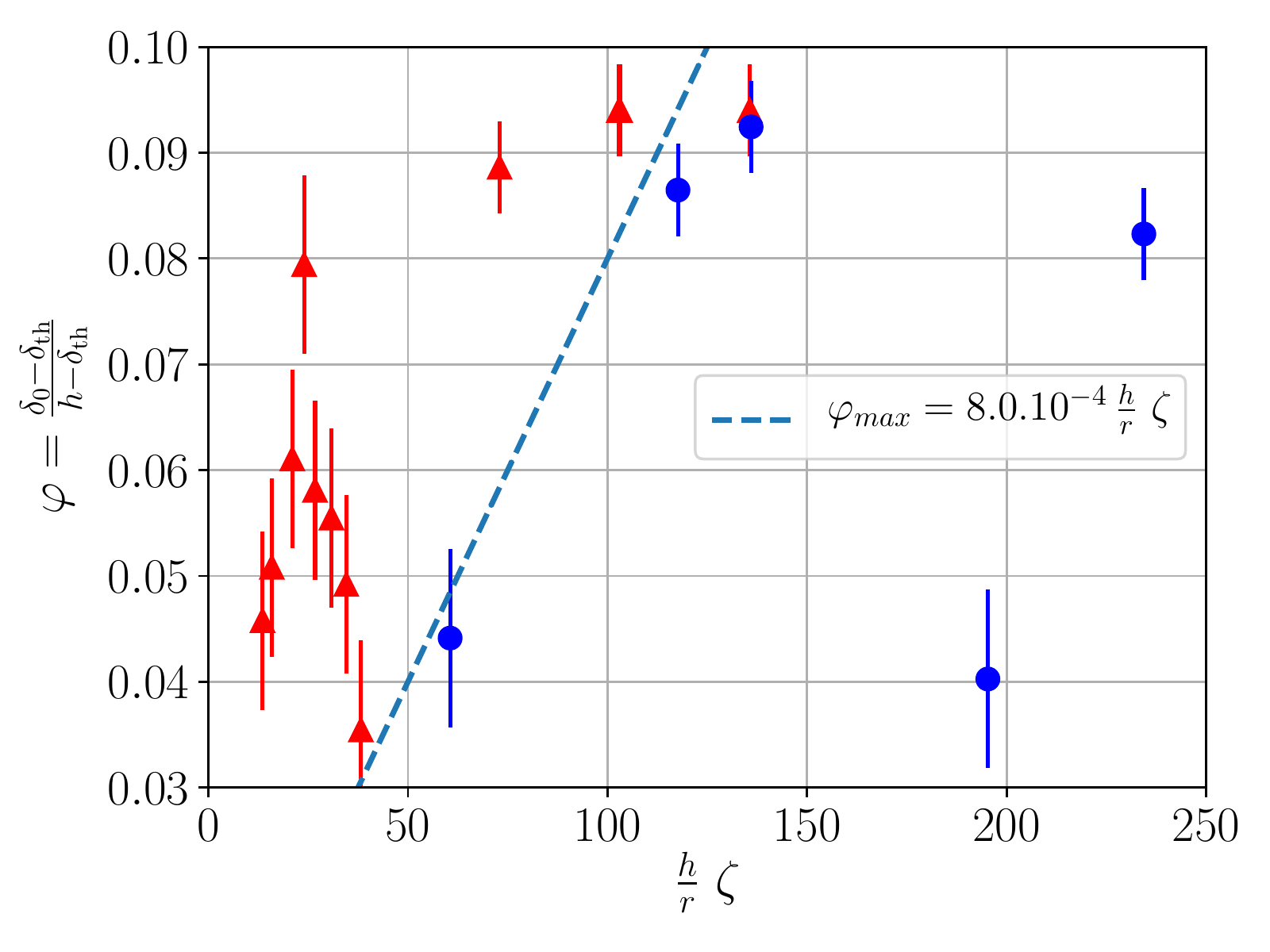}
	\caption{Proxy of particle concentration at steady state for all experiments where particles in suspension are heavier than the fluid. The blue dashed line stands for the transition between the cumulate regime, and the absolute suspension one, based our model based on equation (\ref{eq:431}).}
	\label{fig:11bis}
\end{figure}

	\indent We now illustrate a way to draw a parallel between the two-phase flow framework and the above energy balance approach developed in section \ref{sec:NRJ_balance}. The purpose is to show qualitatively how the efficiency parameter $\epsilon$ could be refined and how it relates to the different regimes that are reached in our experiments. This parameter is supposed to depend on many factors such as fluid and particles densities, the particle size, the concentration and the heat flux involved \citep{Solomatov93a} but this dependency has not yet been clarified. \\
	\indent In a convective system without particles, there is a bulk equilibrium between the total viscous dissipation $\mathcal{V}_{f}$ and the total buoyancy of the fluid $\mathcal{B}_{f}$. In the presence of particles, part of the viscous dissipation corresponds to the work done to maintain the particles in suspension:
	\begin{equation}
		\mathcal{B}_{f}=(1-\epsilon)\mathcal{V}_{f}, \label{eq:423}
	\end{equation}
	with:
	\begin{eqnarray}
		\mathcal{B}_{f}= \iiint_{(V)}\alpha_{f}\rho_{0,f}\theta \mathbf{g}\cdot \mathbf{u}_f\mathrm{d}V. \label{eq:424}
		\end{eqnarray} 
	Combining (\ref{eq:212}) with (\ref{eq:423}) leads to:
	\begin{equation}
		\epsilon \iiint_{(V)}\eta_{f} \mathbf{u}_{f}\cdot\nabla^{2}\mathbf{u}_{f}\mathrm{d}V = \iiint_{(V)}\frac{\mathbf{f}}{1-\phi} \cdot \mathbf{u}_{f}\mathrm{d}V. \label{eq:426}
	\end{equation}
	
	This relation links quantitatively the efficiency parameter to the coupling force between the fluid and particles. In our case, for laminar convection, the coupling force (\ref{eq:214}) can be used and we obtain:
	\begin{equation}
		 \iiint_{(V)}\mathbf{f}\cdot \mathbf{u}_{f}\mathrm{d}V=\iiint_{(V)}\beta \frac{\eta_{f}}{r^{2}}(\mathbf{u}_{f}-\mathbf{u}_{p})\cdot \mathbf{u}_{f}\mathrm{d}V. \label{eq:427}
	\end{equation}
	\indent As discussed earlier, as $St\ll1$, particles are statistically passive tracers. Nevertheless, we observed experimentally the formation of cumulates, so there is a weak component of particle motion that enables settling. In this way, we assume the following decomposition for the particle velocity:
	\begin{equation}
		\mathbf{u}_{p}=\mathbf{u}_{f}+\mathbf{u}_{s}, \label{eq:428}
	\end{equation}
	where $\mathbf{u}_{s}$ stands for the settling velocity, which is considered to be the Stokes velocity, and verifies $||\mathbf{u}_{f}||\gg||\mathbf{u}_{s}||$ in our case. This hypothesis is also corroborated by the experiments done by \cite{Lavorel09}, who showed that particles in turbulent convection settle at a speed that scales with the Stokes velocity. Equation (\ref{eq:427}) thus suggest the (order-of-magnitude) balance:
	\begin{eqnarray}
		\epsilon \eta_{f} \frac{U_{L}^{2}}{h^{2}}\sim \frac{\beta(\phi)}{1-\phi} \frac{\eta_{f}}{r^{2}}U_{s}U_{L}, \label{eq:429}
	\end{eqnarray}
	where the convective velocity scales are $U_{L}\sim \kappa_{f}/h\, Ra_{H}^{3/8}$, $U_{s}\sim\Delta \rho g r^{2}/\eta_{f}$, and where  $\beta(\phi)/(1-\phi)\approx 0.5-1$ in our experiments, which yields:
	\begin{eqnarray}
		\epsilon\sim \frac{h^{2}}{r^{2}}\frac{U_{s}}{U_{L}}\sim \frac{\overline{\Delta \rho} g h^{3}}{\eta_{f} \kappa_{f} }\, Ra_{H}^{-3/8}, \label{eq:430}
	\end{eqnarray}
	which relates the efficiency parameter to the physical properties of the fluid and particles. \\
	\indent To verify this expression in our experiments, we rewrite (\ref{eq:421}) by substituting $\epsilon$ with (\ref{eq:430}) and deduce a new scaling law for the maximal concentration of particles in the suspension:
	\begin{equation}
		\phi_{\mathrm{max}}\sim \frac{h}{r} \overline \zeta. \label{eq:431}
	\end{equation}
	As a consequence, the proxy $\varphi$ of the particle concentration should also follow this law: $\varphi=C_{\varphi} h/r  \overline \zeta$ with $C_{\varphi}$ a constant that is constrained experimentally $C_{\varphi}=8.10^{-4}$ (Figure \ref{fig:11bis}).\\
	\indent To verify qualitatively the consistency of this expression with $\epsilon$, we compare (\ref{eq:421}) and (\ref{eq:431}) to get an expression for the efficient coefficient:
	\begin{equation}
		\epsilon=\frac{C_{\varphi}a}{C_{s}}\, \frac{\overline{\Delta \rho} g h^{3}}{\eta_{f} \kappa_{f} }\, Ra_{H}^{-3/8}. \label{eq:432}
	\end{equation}
	Assuming that particles are packed randomly in the floating lid ($a=0.56$), we get the average value of $\epsilon = 4\pm3 \%$ in our experiments. This estimate is slightly higher than the values reported in the literature but still consistent with them \citep{Solomatov93a,Lavorel09}. However, the main purpose of this physical argument is to show how to reconcile Solomatov's ad-hoc expression with the complete physical framework that governs the two-phase flow and to verify it qualitatively. Experiments where the bulk concentration  of beads  is precisely measured are required to corroborate quantitatively this reasoning.


\subsection{Application to magma oceans}

\subsubsection{Crystal segregation in magma ocean}

\begin{figure}
	\centering
	\includegraphics[width=0.6\textwidth]{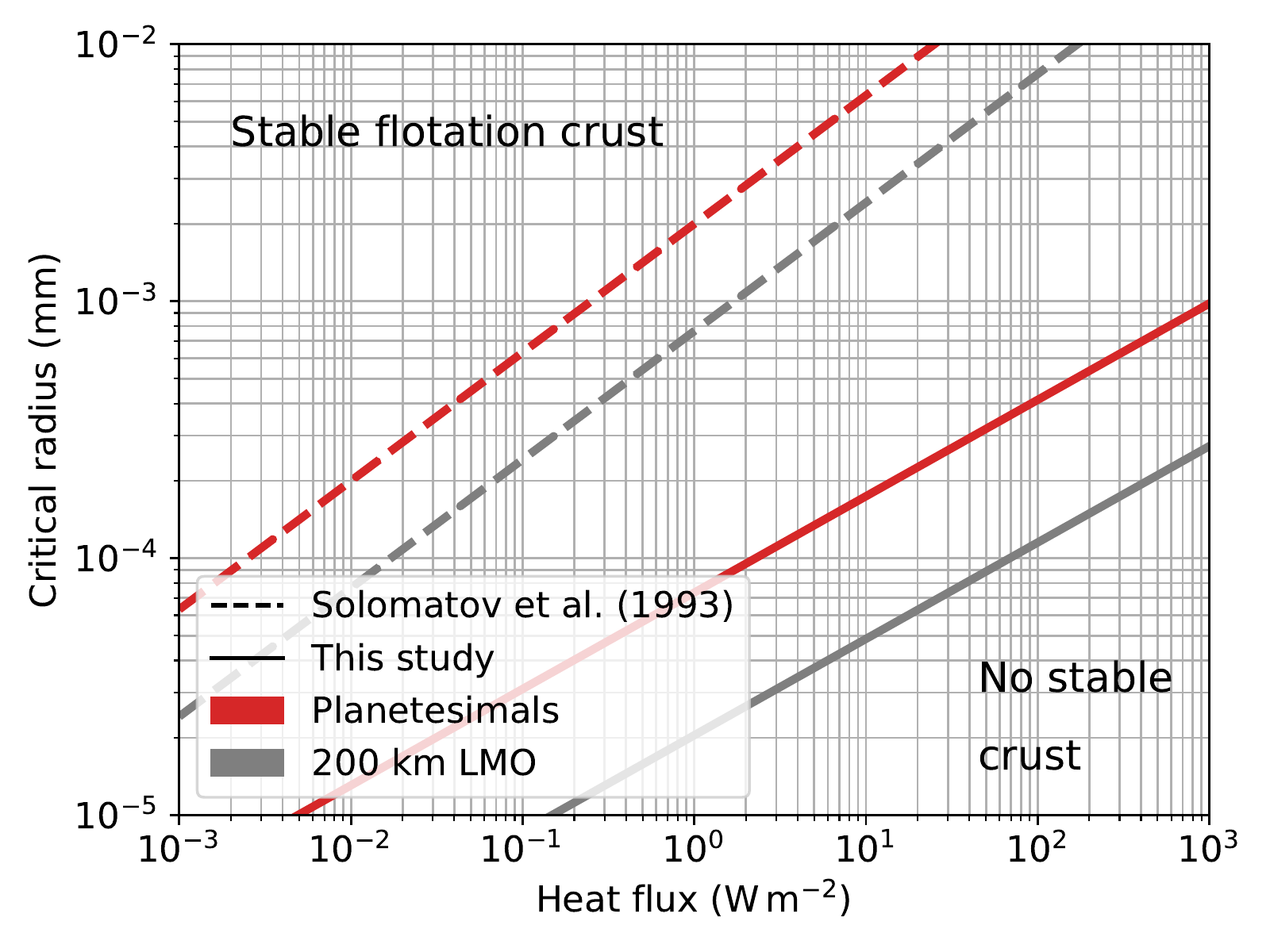}
	\caption{Diagram of flotation versus suspension for two planetary bodies. Each solid line corresponds to the critical radius $r_c$ given by (\ref{eq:CSLaw}). Particles with radius greater than the critical one are allowed to float, otherwise they stay in suspension. Dashed line show the critical radius of crystals predicted by Solomatov's law (\ref{eq:SLaw}).}
	\label{fig:13}
\end{figure}
	\indent Based on the model developed above, we propose an insight into the segregation process of particles in the thermal history of a magma ocean, and especially the flotation of plagioclase. This type of crystals is known to be lighter than the residual liquid from which they form \citep{ElkinsTanton11}. The flotation of plagioclase is the main scenario invoked to explain the formation of the light anorthosite crust of the Moon \citep{Wood70a,Solomon77}.  Convection driven by secular cooling is the relevant regime for convective magma ocean. Secular cooling is mathematically equivalent to internal heating treated in our model \citep[see, e.g.:][]{Krishnamurti68}. \\
	\indent \cite{Deschamps12} demonstrated that scaling laws that describe thermal convection in volumetrically heated cartesian boxes is also valid in spherical geometry, provided we take into account a geometrical factor describing the shell curvature. In this case, the surface heat flux is linked to the secular cooling and the internal heating rate as follows:
	\begin{equation}
		Q_{s}=\frac{R(1-f)}{3}\left(H-\rho c_{p} \frac{\partial \overline T}{\partial t}\right), \label{eq:55}
	\end{equation}
	where $f=(R-h)/R$ is the ratio between the depth of the magma ocean $h$ and the planetary radius $R$. Further details can be found in Appendix \ref{App2}.\\
\indent $Q_{s}$ ranges from $10^{6}\, \mathrm{W\,m^{-2}}$ for molten planetary bodies that release heat by radiation to space \citep{ElkinsTanton11, Massol16}, down to $10^{-3}-10^{-2}\, \mathrm{W\,m^{-2}}$ when the flotation crust is present \citep{Maurice20}. All physical quantities used in the model are summarized in table \ref{tab:5}.\\
	\indent We deduce the critical particle radius enabling crystal flotation for two types of magma ocean: a 200 km deep shallow lunar magma ocean ($R=1737\ \mathrm{km}$, $f=0.88$, $g=1.6\, \rm{m\, s^{-2}}$) and a fully molten planetesimal ($R=300\ \mathrm{km}$, $f=0.42$, $g=0.23\, \rm{m\, s^{-2}}$). This critical radius $r_c$ is calculated from the critical Shields number (\ref{eq:416}): 
	\begin{equation}
		r_c=\frac{\kappa \eta}{h\Delta \rho g \zeta_c}\, \left(\frac{\alpha \rho g Q_s h^4}{\eta \kappa \lambda}\right)^{3/8}. \label{eq:CSLaw}
	\end{equation}
	Results are displayed in figure \ref{fig:13}. As crystal size during crystallization is estimated in the range $r=0.1-10\, \mathrm{mm}$ \citep{Solomatov00}, we deduce that crystals float during magma ocean cooling. Magma ocean episodes enable efficient crystal segregation. This segregation is less efficient for smaller bodies such as planetesimals (with $R\le300\, \mathrm{km}$), suggesting that crystallization history might be more complex in these cases.We compared our results to those displayed in \cite{ET12}  based on the law of \cite{Solomatov93a}:
	\begin{equation}
		r_c=\frac{1}{2\Delta \rho g}\, \left(\frac{0.1 \eta \alpha g Q_s}{c_p}\right)^{1/2}, \label{eq:SLaw}
	\end{equation}
	with $\eta=0.1\, \rm{Pa\, s}$.  (\ref{eq:SLaw}) was considered by these authors to be applicable in both laminar and turbulent experiments but established for system heated from below. Here we propose a model for convective systems driven by secular cooling. Our model enables smaller particles to participate to crustal formation.\\

\begin{table}
	\centering
	\begin{tabular}{ c c c}
		\hline
		Physical properties 				& Value 		& Units\\
		\hline
		Melt density $\rho$			& 2800		& $\rm{kg\ m^{-3}}$\\
		Density difference $\Delta \rho$		& 100		& $\rm{kg\ m^{-3}}$\\
		Viscosity $\eta$				& 100		& $\rm{Pa\ s}$\\
		Specific heat $c_{p}$ 			& 800		& $\rm{J\ K^{-1}\ kg^{-1}}$\\
		Thermal conductivity $\lambda$	& 4			& $\rm{W\ m^{-1}\ K^{-1}}$\\
		Thermal expansion $\alpha$		& $3.10^{-5}$	& $\rm{K^{-1}}$\\
		\hline
	\end{tabular}
	\caption{Physical parameters used in the model. }
	\label{tab:5}
\end{table}


\begin{figure}
	\centering
	\includegraphics[width=0.6\textwidth]{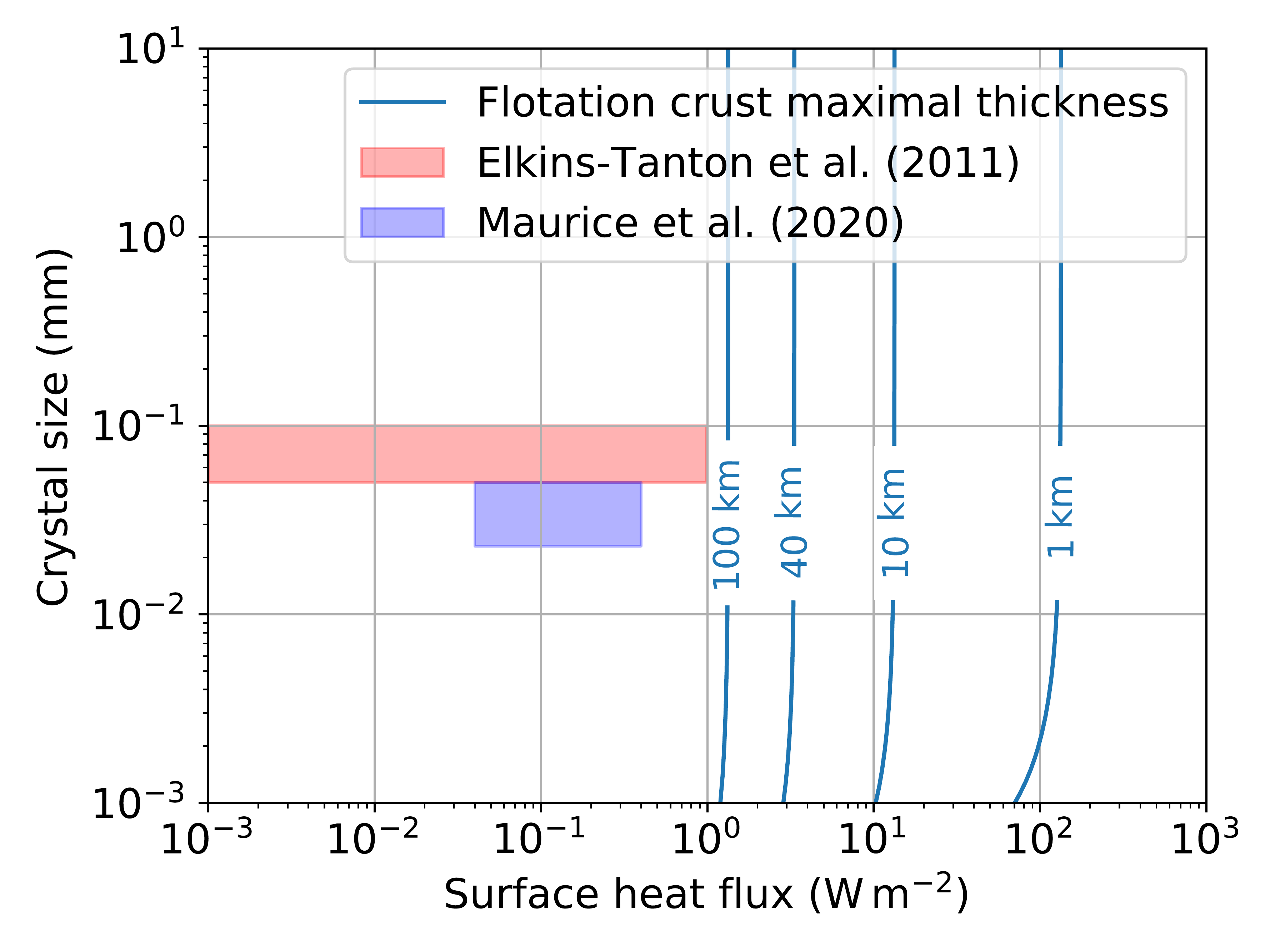}
	\caption{Evolution of the maximal thickness of the anorthosite crust formed by plagioclase flotation above the lunar magma ocean as a function of the crystal size and the surface heat flux. We represent also the range of surface heat flux that are simulated by two thermal models \citep{ElkinsTanton11, Maurice20}.}
	\label{fig:15}
\end{figure}

	\subsubsection{Crustal thickness and flotation efficiency}
	
	\indent Our model not only provides criteria for the formation of stable crystal reservoirs, floating crust or cumulates, but can also be used to quantify the partitionning of the crystals between the suspension and these deposits. We illustrate how to quantify this partition of crystals between the deposits and the convective bulk in the  particular case of the lunar magma ocean (LMO).\\
	\indent In the literature, it is widely assumed that all plagioclase crystals settle instantaneously once they nucleate in the LMO \citep{ElkinsTanton11,Maurice20}, which yields a system composed of a flotation crust growing above a purely liquid magma ocean. The main results of these two studies are indicated in the two highlighted bands in figure \ref{fig:15}. This hypothesis was relaxed using the flotation efficiency which is the ratio between the amount of crystals that float and the total volume of crystals that nucleate in the system. This parameter can be estimated from field data. For instance, \cite{Namur11} determined the value for plagioclase flotation by studying the Sept Iles layered intrusion, and they found an efficiency of 50\%. They assumed that this flotation efficiency would be the same in the case of the LMO, and they use it to constrain the amount of plagioclase that composed the anorthosite crust and its thickness. \cite{Charlier18} used the crustal thickness measured by spacecraft gravity data  \citep{Wieczorek12} and a flotation efficiency in the range between 40 and 100\% in order to constrain the thickness of the LMO. \\
	\indent  This ad-hoc parameter can be discussed in the light of our model. According to the present study, if the initial volume of crystals is smaller than the critical volume that can be sustained at the surface, all crystals form a flotation crust. If the crust has reached its maximum thickness, the crystals ``in excess'', i.e. that can not be incorporated in the crust, remain in suspension. This is a fundamentally different way of thinking which links crystal flotation to physical parameters.\\
	\indent We consider that the LMO is a shallow shell corresponding to 20 vol\% of the total volume of the Moon. This high level of solidification is necessary to trigger plagioclase crystallization \citep{ElkinsTanton11}.  The evolution of the maximal thickness sustainable at the surface of the LMO is displayed in figure \ref{fig:15}. The crystal size has almost no influence on the steady state crustal thickness except on the low limit. This is due to the sub-critical value of the Shields number, which is essentially constrained by the low value of the melt viscosity. In order to form a crust with a thickness of 40 km corresponding to current estimates of the lunar crust, our model predicts that the surface heat flux should be below $3\, \mathrm{W\ m^{-2}}$. In the cases of \cite{ElkinsTanton11} and \cite{Maurice20}, 100\% flotation efficiency would imply a thicker crust. Our model adds some physical constraints on the LMO evolution in general, by coupling quantitatively the segregation of crystals to the thermal history. \\
	\indent We emphasize that our reasoning is dealing with stability of the crust and its steady state. The predicted crust thickness is reached after a time-dependent deposition process which is not discussed in this paper. It requires a complete description of the transient regime of the flotation crust.


	\subsubsection{Limitation of the model}

\indent We aim to illustrate how the criterion developed in this paper can be applied to natural systems and how they can provide new constraints on conditions under which a deposit can form or not. Of course, our illustration is a crude approximation of the complexity of magma ocean. We use simplified hypotheses, but the framework we propose here is robust enough to enable refinements that can be added at will in order to include complexity.\\
\indent Our first strong assumption is the bulk crystallization that was abundantly considered in the literature as a common end-member \citep{Solomatov00,ElkinsTanton11}. This is justified by the nature of convection driven by secular cooling, which implies that crystals nucleate within the cold downwellings that end up in an isentropic convective bulk. For small-size bodies, this is equivalent to an isothermal interior.\\
\indent We also assume that crystals are non-cohesive, as in experiments. This hypothesis is compatible with batch crystallization where crystals nucleates in the convective bulk. This regime of crystallization is consistent  with magma oceans, but may not be for smaller magmatic reservoirs such as magma chambers or lava lakes. In these cases, crystals may mainly form at the boundary of the system, directly where they are supposed to settle. This adds a complexity to the system as cohesion between crystals becomes important. One way to take cohesion into account in the present model may be by increasing the erosion threshold $\zeta_c$. The stronger the interaction between crystals are, the higher the value of $\zeta_c$. \\
\indent In addition, we did not consider in detail the effect of crystallization series on the evolution of composition and its consequences on all the physical parameters such as the number of crystals, crystal size and liquidus and solidus temperatures. We expect that in limit of small crystal fractions, our results would still apply, but the transient behavior of crystallizing systems needs to be studied specifically.\\
\indent Moreover, our model does not take into account the influence of pressure on all the physical parameters. This hypothesis is consistent with small planetary bodies such as planetesimals, or with the latest stages of shallow magma oceans, but it is less justified for larger bodies such as the Earth. In the latter case, pressure would affect the whole thermodynamics, as it influences the geotherm, the solidus and liquidus of crystals, the viscosity of the liquid magma, the formation of crystals, etc. In these cases, our stability criterion based on the Shields number is still valid, but the threshold value and the scaling laws that are involved should be adapted accordingly.\\
\indent Furthermore, we described the system by adopting stability criterion as did Solomatov. But magma oceans are evolving systems where temperature varies a lot during the thermal history. Hence, describing their complex behavior over time thanks to a stationary criterion is a strong assumption that has to be relaxed. Complete thermal model that takes into account the kinetics of sedimentation and erosion is required to deal with suspension stability in magma oceans. Erosion/deposition mechanism commonly used in geomorphology \citep{Charru04,Lajeunesse10} could be further revisited in terms of crystal erosion and sedimentation in magmatic reservoirs.\\
\indent We considered here the high Prandtl limit where inertia is negligible compared to viscous forces. Nonetheless, magma oceans have experienced turbulent episodes \citep{Solomatov00} that imply high Rayleigh numbers that are not yet attainable experimentally or numerically. This seems to be all the more true for large bodies. In this case, scaling laws used for the thermal state have to be adapted. For instance, \cite{Kraichnan62} expressed the different scaling law for turbulent flow, depending on the Prandtl number. It would lead to other scaling laws for the shear stress and thus other expressions for the Shields number. But fundamentally, the bases of our approach would remain true even in the turbulent case.


\section{Conclusion}

\indent Particles sheared in a convective fluid can form deposits, floating crusts or cumulates, depending on the sign of their buoyancy. Using laboratory experiments, we proposed a model that estimates the partitioning of particles in such systems at high Prandtl number and for Rayleigh-Roberts number up to $10^{9}$. Our model is based on the estimation of one dimensionless parameter that encapsulates the main physical ingredients required to describe the system: the generalized Shields number $\zeta$ that compares the buoyancy of particles to the shear stress generated by convection. The value of this parameter quantifies both the thickness of the flotation crust, and the maximal concentration of crystals that can be sustained in suspension. The volume of particles that forms a cumulate can be deduced from these two  pieces of information by a mass budget. This unifying framework can be adapted to understand transient episodes of the thermal history of natural systems driven by secular cooling and/or internal heating. In order to do so we need to complete the model by the detailed description of the dynamics of erosion and deposition, and in particular of the characteristic timescales that are involved. These timescales have to be compared to the convective system cooling timescale.   Further studies are needed to completely understand the feedback existing between these intricate timescales. This is a necessary step in order to explain the wealth of the observed structures in planetary bodies and in their vestiges that arrive to us as meteorites.

\section*{Acknowledgements}
\indent This paper is part of Cyril Sturtz's PhD thesis (Universit\'e de Paris, Institut de Physique du Globe de Paris). The authors thank Pr. Chomaz and two anonymous reviewers for their fruitful comments that improved the paper. This study contributes to the IdEx Universit\'e de Paris ANR-18-IDEX-0001. This work was supported by the Programme National de Plan\'etologie (PNP) of CNRS/INSU, co-funded by CNES

\section*{Supplementary materials}
\indent Supplementary materials are available at ... .

\section*{Declaration of interest}
\indent The authors report no conflict of interest.

\appendix


\section{Reynolds numbers and dynamical boundary layer}
\label{App1}

\begin{figure}
	\centering
	\includegraphics[width=0.95\textwidth]{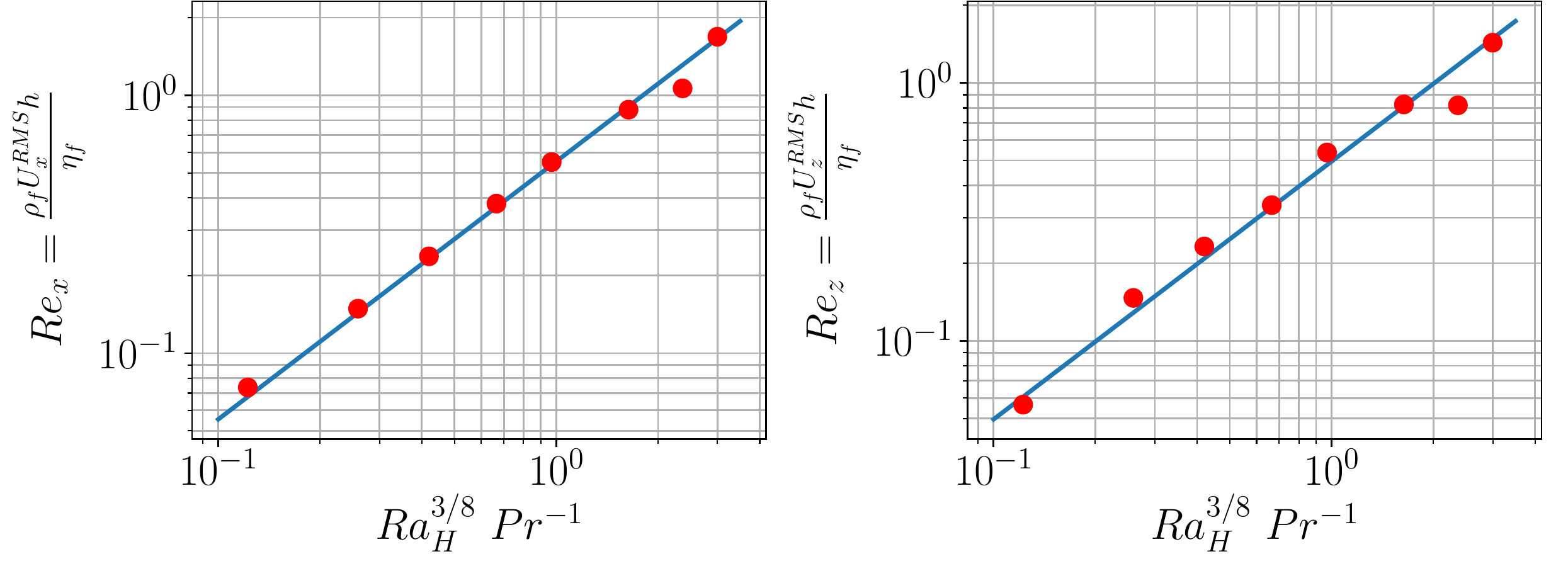}
	\caption{Scaling law for the horizontal (left) and vertical (right) Reynolds numbers in internally heated convective systems.}
	\label{fig:16}
\end{figure}

\begin{table}
	\centering
	\begin{tabular}{c c c}
		\hline
		Variable 								& Fixed exponent\\
		\hline
		$Re_{x}$								& $0.55\ Ra_{H}^{3/8}\ Pr^{-1}$\\
		$Re_{z}$								&  $0.50\ Ra_{H}^{3/8}\ Pr^{-1}$\\
		 \hline
	\end{tabular}
	\caption{Parameters of power laws determined experimentally for the vertical and horizontal Reynolds numbers.}
	\label{tab:Re}
\end{table}

\indent With rigid boundary condition, one can define the dynamical boundary layer (DBL) as the region where the velocity tends to zero. The thickness of the DBL can be expressed from the balance between the convective term and the diffusion term in the conservation of momentum equation \citep[e.g:][pp.128]{JaupartMareschal}:
		\begin{equation}
			\rho_{f}\frac{U_{L}^{2}}{h}\sim \eta \frac{U_{L}}{\delta_{v}^{2}}.
		\end{equation}
which leads to $\delta_{v}\sim h\ Re^{-1/2}$ where $Re=\rho_{f}hU_{L}/\eta_{f}$ is the Reynolds number. This relation justifies quantitatively that the dynamical boundary layer occupies the entire reservoir in the laminar limit, which is the regime reached in our experiments (figure \ref{fig:16}). 
\indent The scaling law that characterizes velocity in internally heated convective systems can be rewritten in terms of Reynolds number $Re$ using (\ref{eq:412}):
\begin{equation}
	Re\sim Ra_{H}^{3/8}\ Pr^{-1},
\end{equation}
which has been verified experimentally as illustrated in figure \ref{fig:16}. The pre-factors for horizontal and vertical Reynolds number are given in table \ref{tab:Re}.


\section{Secular cooling and internal heating}
\label{App2}

\indent In geophysical systems evolving over long periods of time, the thermal state is transient, so one can argue that our reasoning is not applicable as scaling laws that are used here hold only at steady state. Here, we will show that it is possible to adapt them to describe the transient state. \\
\indent First, we treat the floating lid as a homogeneous layer where Fourier's law can be applied. In the transient state, the heat flux departs from (\ref{eq:41}), which is valid only in steady-state conditions. Hence, we can define the diffusive time scale in the floating lid $\tau_{\mathrm{diff}}=\delta^{2}/\overline \kappa$, with $\overline \kappa=\overline \lambda/ \overline \rho\, \overline c_{p}$ the thermal diffusivity of the floating lid where $\overline \lambda$, $\overline \rho$ and $\overline c_{p}$ are respectively its average thermal conductivity, density and specific heat. This time scale is compared to the time scale related to the evolution of the thermal state of a convective fluid heated from inside. The latter is given by: $\tau_{\mathrm{conv}}=h^{2}/\kappa_{f}\, Ra_H^{-1/4}$ \citep{Limare19}. Thus, the criterion that ensures that the lid's thermal state evolves quasi-statically is based on the ratio:
\begin{equation}
	R_{\tau}=\frac{\tau_{\mathrm{diff}}}{\tau_{\mathrm{conv}}}=\frac{\overline \kappa}{\kappa_{f}}\, \left(\frac{\delta}{h}\right)^{2}\, Ra_{H}^{1/4}. \label{eq:51}
\end{equation}
If $R_{\tau}\ll1$, the lid reaches its steady state quickly and (\ref{eq:41}) holds true. In our experiments, $R_{\tau}\approx 0.1-0.5$, so the approximation is justified. In geophysical systems, the approximation holds true if $Ra_{H}$ is moderate, and/or if the floating lid is thin compared to the depth of the convective mantle. For instance, for lids that represent $\delta/R=1\%$ equivalent to the ratio between the lithosphere thickness and the Earth's radius, the quasi-static approximation is correct for $Ra_{H}$ up to $10^{12}$, which is relevant for geophysical applications.\\
\indent Second, the model is based on scaling laws that describe the thermal state of convective systems heated internally. In the transient state, the secular cooling term can be considered as an additional source of internal heat. We define the modified rate of internal heat generation as:
\begin{equation}
	H^{*}=H-\rho_{f} c_{p,f}\frac{\partial T}{\partial t}. \label{eq:52}
\end{equation}
In this way, the Rayleigh-Roberts number is also modified:
\begin{equation}
	Ra_{H}^{*}=\frac{\alpha_{f} \rho_{0,f} g H^{*}h^{5}}{\eta_{f} \kappa_{f} \lambda_{f}}.  \label{eq:53}
\end{equation}
Thus, the drop of temperature across the TBL becomes:
\begin{eqnarray}
	\Delta T_{TBL}&=&C_{T}^{*}\, \frac{H^{*}h^{2}}{\lambda_{f}}\, (Ra_{H}^{*})^{-1/4}.  \label{eq:54}
\end{eqnarray}
\cite{Limare19X} showed that this model describes well the transient state in experimental convective systems. The authors measured $C_{T}^{*}=3.58\pm0.15$ similar to the value retrieved from experiments in homogeneous, steady state conditions. This model was further used to describe the thermal evolution of parent bodies of iron meteorites \citep{Kaminski20}. It shows that convective systems where convection is mainly driven by secular cooling can also be treated in the framework presented in this paper.\\
\indent Consequently, we are able to estimate the evolution of the Shields number from the thermal history of such systems, enabling the study of the formation of floating crust and/or deposits.

\bibliographystyle{jfm}
\bibliography{biblio}

\end{document}


\maketitle


\section{Experimental apparatus}

\begin{figure}[h]
	\centering
	\includegraphics[width=0.6\textwidth]{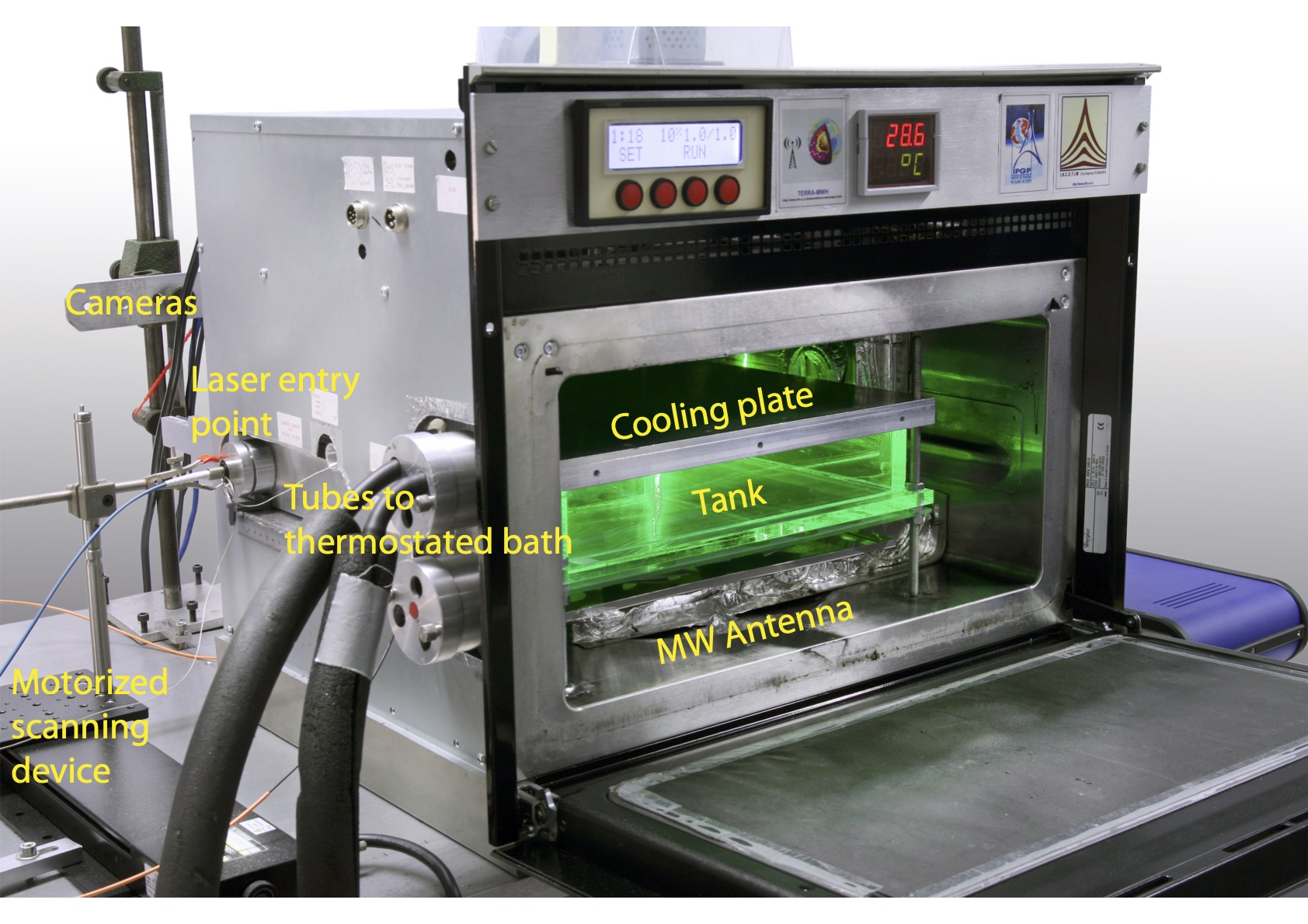}
	\caption{Picture of the setup used in this study (taken from \cite{Limare19}).}
\end{figure}

\section{Properties of fluids and beads -- methods}

\subsection{Fluid}

\indent The fluid used in this study is a mixture of 44wt\% glycerol and 56wt\% ethylene glycol. Density and thermal expansion have been determined thanks to a DMA 5000 Anton Paar densimeter. Thermal properties, such as thermal diffusivity $\kappa_{f}$, thermal conductivity $\lambda_{f}$ and specific heat $c_{p,f}$, have been determined thanks to a photothermal method described by \cite{Dadarlat09}. The viscosity $\eta_{f}$ and its dependance on temperature have been monitored thanks to an RS600 Anton Paar rheometer. The refractive index $n_{f}$ and its temperature dependence have been obtained thanks to the Abbemat 350 Anton Paar refractometer. These properties are shown in table \ref{tab:ParamsExp}  and their tempertaure dependence is plotted in figure \ref{fig:ExpParams} $(a)$, $(b)$ and $(c)$.

\subsection{Beads}
\begin{figure}
	\centering
	\includegraphics[width=0.45\textwidth]{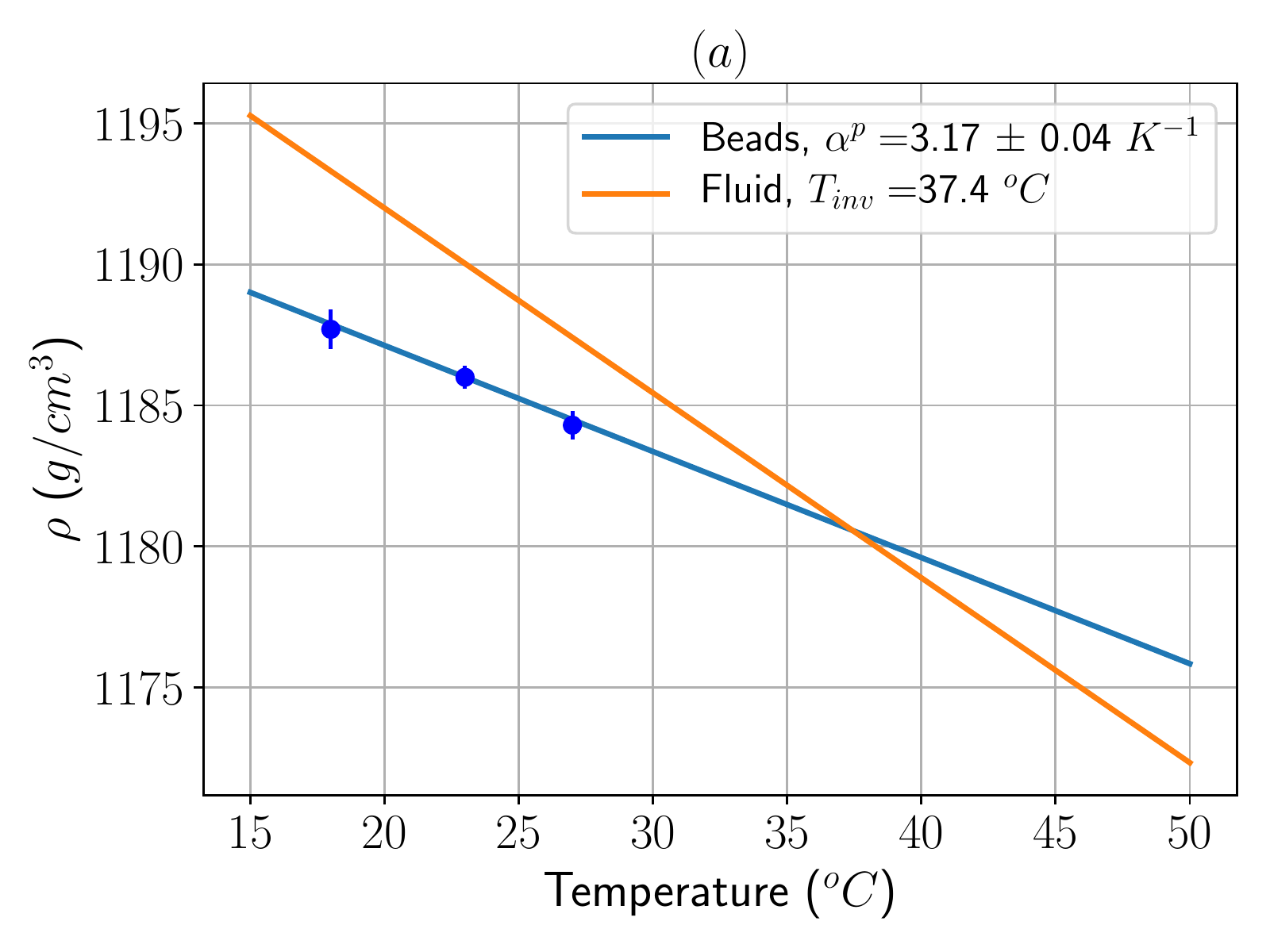}
	\includegraphics[width=0.45\textwidth]{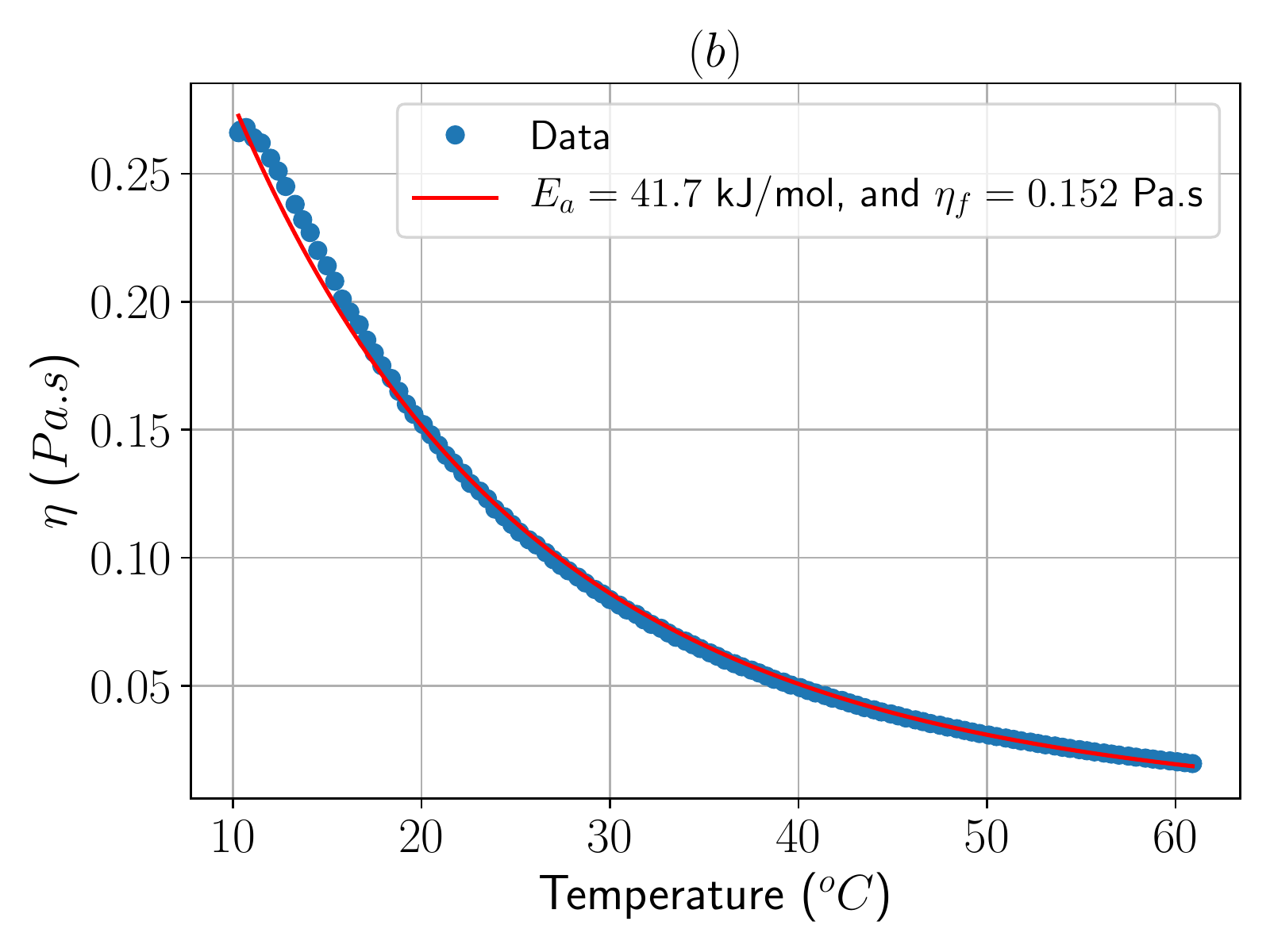}
	\includegraphics[width=0.45\textwidth]{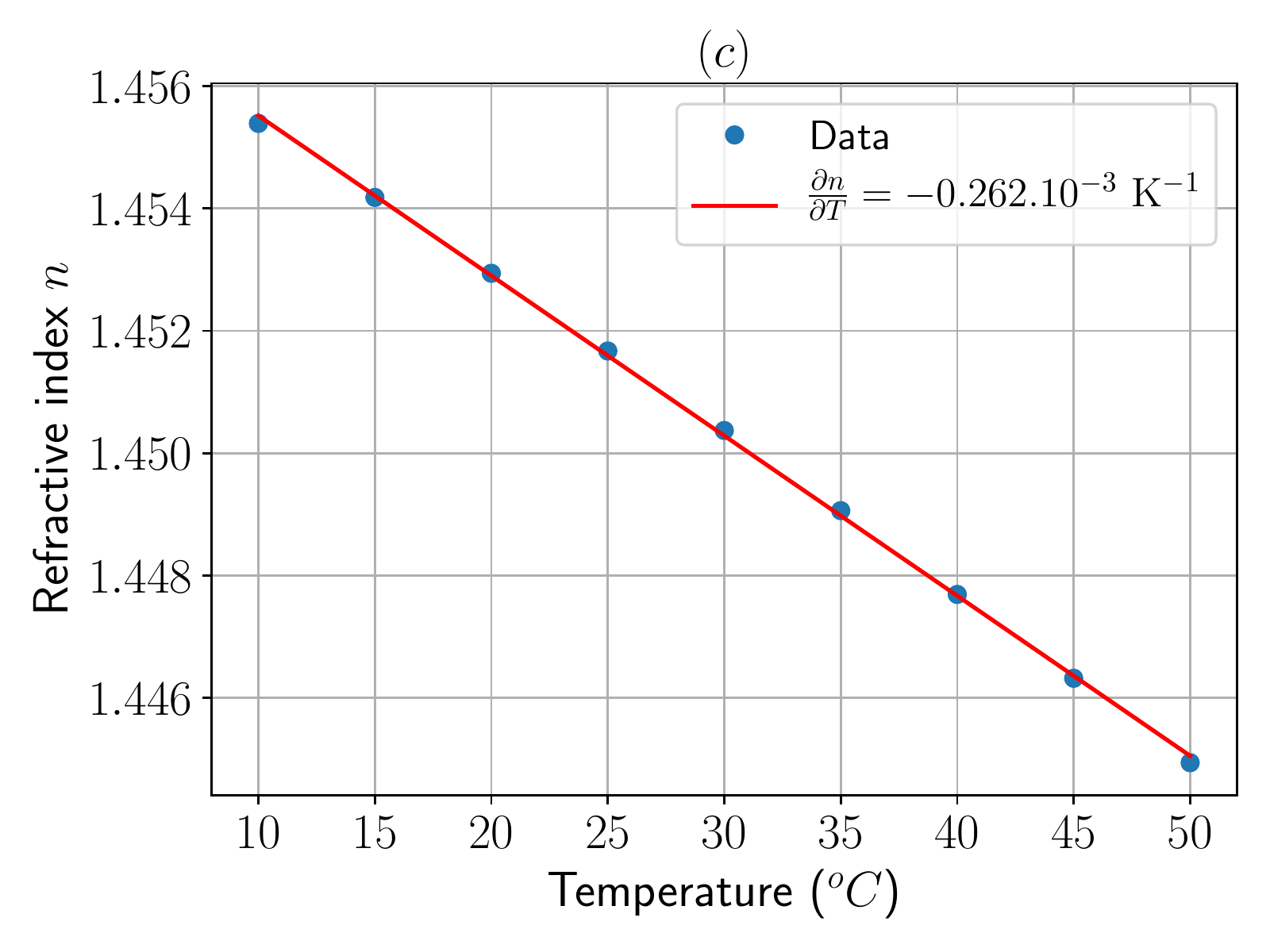}
	\includegraphics[width=0.45\textwidth]{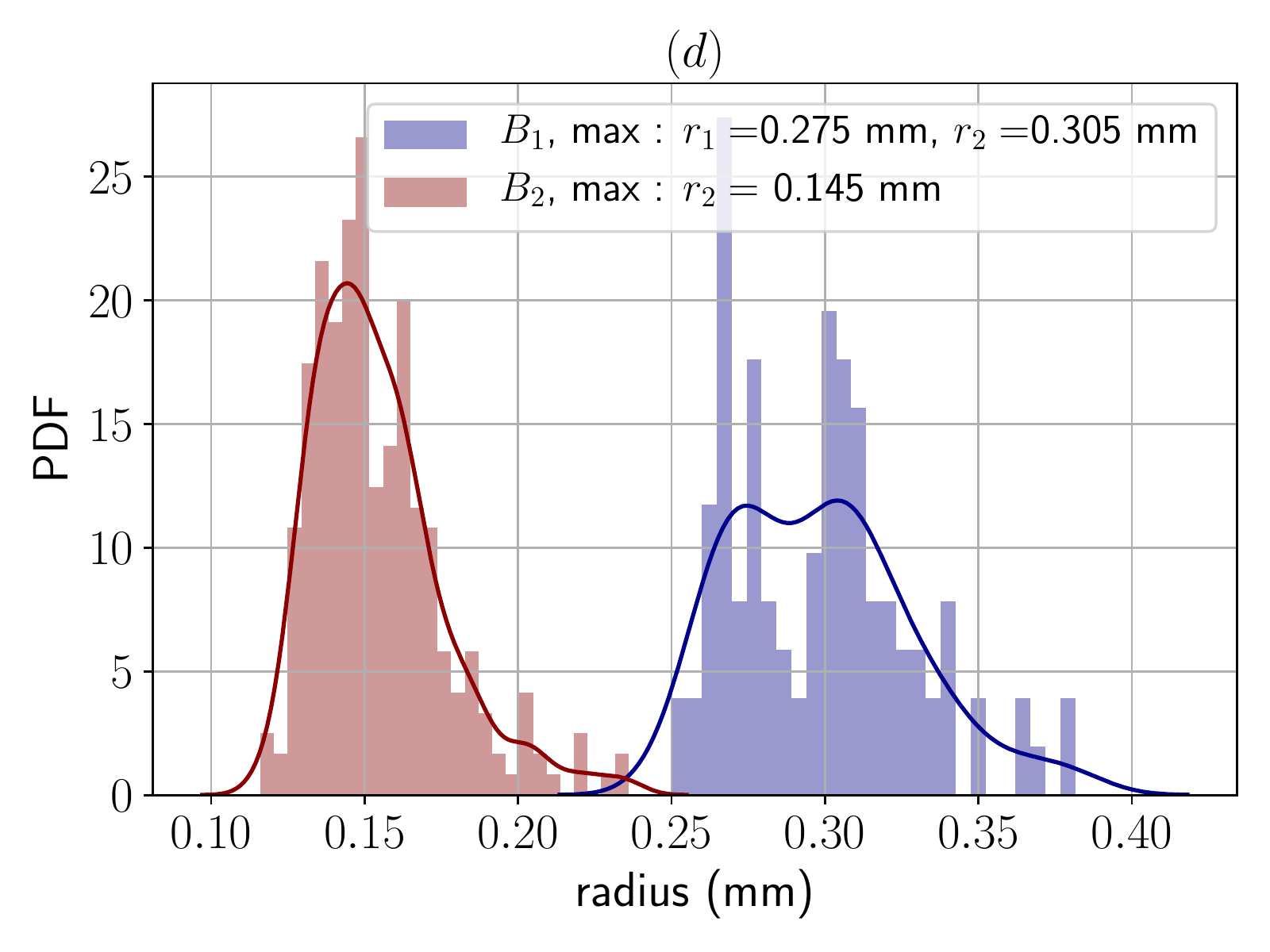}
	\caption{Some properties of the fluid and beads that has been measured. $(a)$: fluid and beads density. Blue dots represent experiments using thermalised stratifications and are described in the text. $(b)$: Fluid viscosity as a function of the temperature. The fit represents an \textsc{Arrhenius} law: $\eta(T)=\eta_{f}\exp\left[\dfrac{E_{a}}{R}\left(\dfrac{1}{T}-\dfrac{1}{T_{0}}\right)\right]$ with $T_{0}=20^{o}C$. $(c)$: fluid refractive index as a function of the temperature. $(d)$: beads radius distribution.}
	\label{fig:ExpParams}
\end{figure}

\begin{figure}
	\centering
	\includegraphics[width=0.9\textwidth]{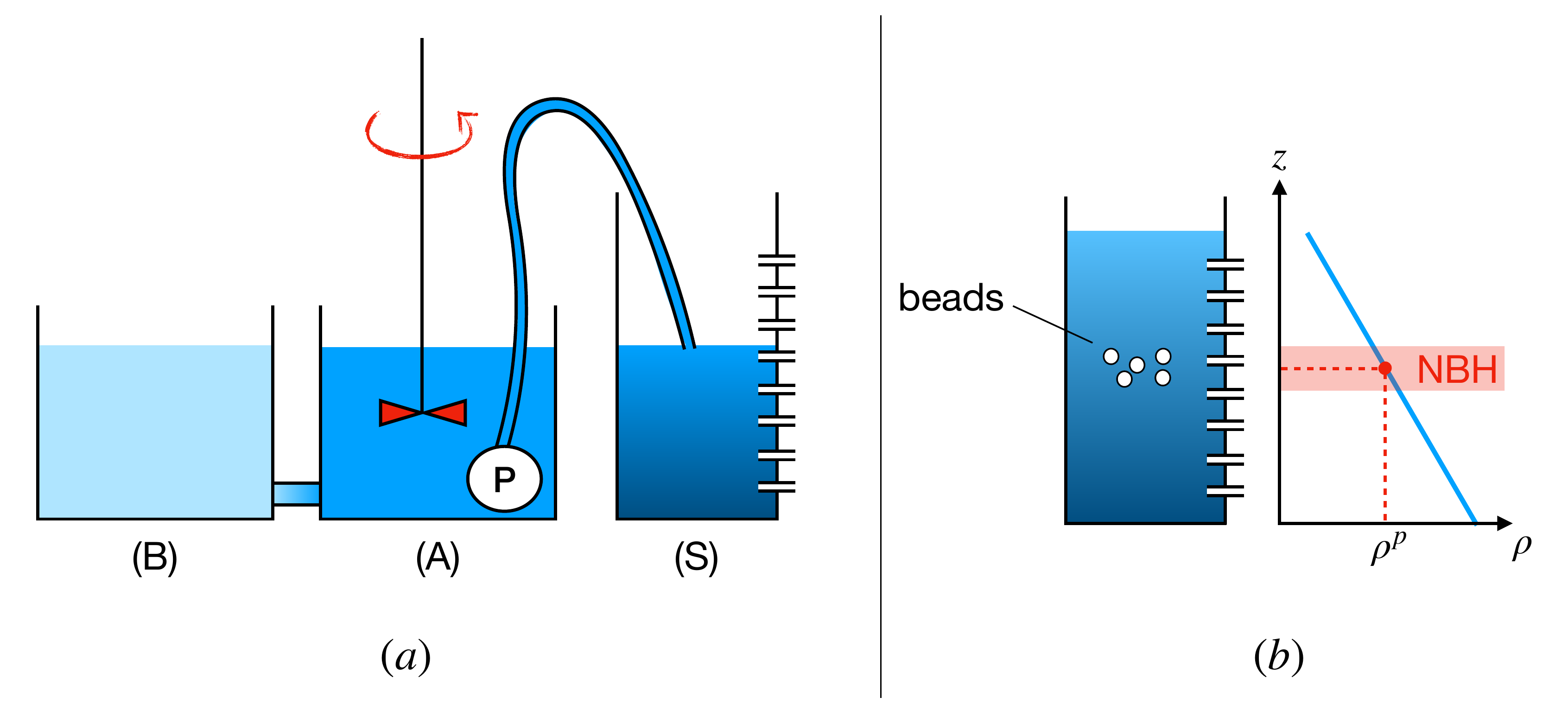}
	\caption{$(a)$: Setup to create a stratified environment (S), using the double-bucket method. The density stratification created in (S) is measured by sampling the liquid in the tank at different heights, as represented at the right wall of (S). $(b)$: scheme showing the measurement of beads density by the determination of their neutral buoyancy height (NBH).}
	\label{fig:strat}
\end{figure}

\indent As beads are small solid spheres, the measure of previous properties is much more complicated. To measure the density, we introduced the beads in a tank containing a stratified fluid. This density stratification  is achieved by the double bucket method described by \cite{Fortuin60} and illustrated in figure \ref{fig:strat}. The buckets contain two mixtures of different proportions of water and glycerol having different densities, called (A) and (B) for the high and low value respectively. As the heavy mixture (A) is pumped into the working tank (S), the lighter one (B) is introduced in (A) thanks to a basal linking tube, reducing the density of (A) and, hence, the density of the fluid deposed in the tank (S). Step by step, as the pumping is going on, the density of  fluid introduced in (S) becomes less important. The stratification so created is sampled at different heights, and controlled apart, thanks to the DMA 5000 Anton Paar densimeter. Beads are introduced in the stratified liquid, and they stop at a level corresponding to their neutral buoyancy heigh (NBH). The determination of this height is a way to measure their density.\\
\indent In order to measure the thermal expansion $\alpha_{p}$, this type of set up has been carried on for different thermostated environments. Achieving a stable and isothermal stratified environment is extremely complicated, because even a difference of a few degrees between the wall of the tank and the bulk might trigger convection, and thus, destroy the stratification. In order to prevent this effect, the fluids (A) and (B) have been prepared in a thermostated room, whose temperature can be tuned within a certain range (between 18 and 27$^{o}$C). In this way, we get three measurements of density at three different temperatures (Figure \ref{fig:ExpParams} $(a)$), and assuming a constant thermal expansion, we get the following value for beads: $\alpha_{p}=3.17.10^{-4}\ \mathrm{K^{-1}}$.\\
\indent We work with two families of beads of different radius. Their size has been determined by image analysis of a sample of  hundreds of beads of each family. The size distribution is plotted in Figure \ref{fig:ExpParams} $(d)$, and it leads to the estimation of the averaged radius: $r_{1}=290\ \mathrm{\mu m}$ and $r_{2}=145 \ \mathrm{\mu m}$.\\
\indent All the other thermal properties are taken in \cite{HandbookPolymer}: $\lambda_{p}=0.21\ \mathrm{W/m/K}$, $c_{p,p}=1765\ \mathrm{J/kg/K}$ and thus $\kappa_{p}=1.10^{-7}\ \mathrm{m^{2}/s}$. Concerning the refractive index, the handbook exposes a wide range of values, from $1.445$ for bead to $1.56$ (\cite{HandbookPolymer}, p.830). As we noticed a strong index mismatch in experiments, the refractive index of the beads is likely to be closer to the upper bound of this range.

\begin{table}
	\centering
	\begin{tabular}{c c c }
	\hline
	Properties 					& Symbol 			&Value\\
	\hline
	Fluid density ($20^{o}C$)			& $\rho_{0,p}$ 		& $1192$ $\mathrm{kg.m^{-3}}$\\   
	Beads density ($20^{o}C$)		& $\rho_{0,p}$		& $1187$ $\mathrm{kg.m^{-3}}$\\
	Fluid thermal expansion 			& $\alpha_{f}$		& $5.5\ 10^{-4}$ $\mathrm{K^{-1}}$\\
	Beads thermal expansion 			& $\alpha_{p}$		& $3.2\ 10^{-4}$ $\mathrm{K^{-1}}$\\
	Fluid viscosity ($20^{o}C$)		& $\eta_{f}$		& $0.151$ Pa.s\\
	Activation energy 				& $E_{a}$			& $41.7$ $\mathrm{kJ.mol^{-1}}$\\
	Fluid thermal diffusion coefficient 	& $\kappa_{f}$		& $9.1\ 10^{-8}$ $\mathrm{m^{2}.s^{-1}}$\\
	Beads thermal diffusion coefficient(*) & $\kappa_{p}$	& $1.\ 10^{-7}$ $\mathrm{m^{2}.s^{-1}}$\\
	Fluid thermal conductivity 			& $\lambda_{f}$	& $0.276$ $\mathrm{W.m^{-1}.K^{-1} }$\\
	Beads thermal conductivity(*) 		& $\lambda_{p}$	& $0.21$ $\mathrm{W.m^{-1}.K^{-1} }$\\
	\hline
	\end{tabular}
	\caption{Main physical properties of the fluid and beads. The activation energy corresponds to the fit of the variation of viscosity with an Arrhenius law. Properties are all measured, except those marked with (*) which are taken from \cite{HandbookPolymer}. }
	\label{tab:ParamsExp}
\end{table}

\section{How to introduce beads without bubbles?}
\indent Achieving an appropriate experimental setup that involves a fluid bearing beads is actually challenging. Especially, we have to avoid introducing air bubbles in the mush, as it might induce unwanted surface tension effects, and, thus, increase cohesion between particles. \\
\indent To do so, we systematically applied the following protocol. First, the working tank is filled with the fluid. Then, we prepared a mixture of fluid and beads in the home-made stir-machine -- see Figure \ref{fig:stir} (a) and (b). The stir-machine includes a vessel and a rotor that is composed of an horizontal grid aiming to maintain the beads in the lower part of the stir-machine, and let the bubbles escape. This protocol is done at room temperature, therefore beads float and form a lid against the grid. In order to prevent settling and to maintain the beads in suspension, three blades are also attached to the rotating grid. Degassing the mixture consists in stirring the fluid with beads, and letting it quiescent in order to let the bubbles escape. This procedure is repeated several times, until no more bubbles escaped from the separation grid. Then, the stir-machine is connected to the working tank. Isostatic pressure enables the beads to go from the stir-machine to the tank. The flow has to be controlled in order not to let particles be entrained and escape the working tank. Once the amount of particles introduced is sufficient, the tank is closed, shaken to homogenize the mixture, and placed into the oven. We waited until particles settle before beginning any experiment. The total beads mass introduced in the tank is measured at the end of the set of experiments, when the tank is drained, the mixture is sieved, and the beads are washed, dried and then weighed. 
\begin{figure}
	\centering
	\includegraphics[width=0.8\textwidth]{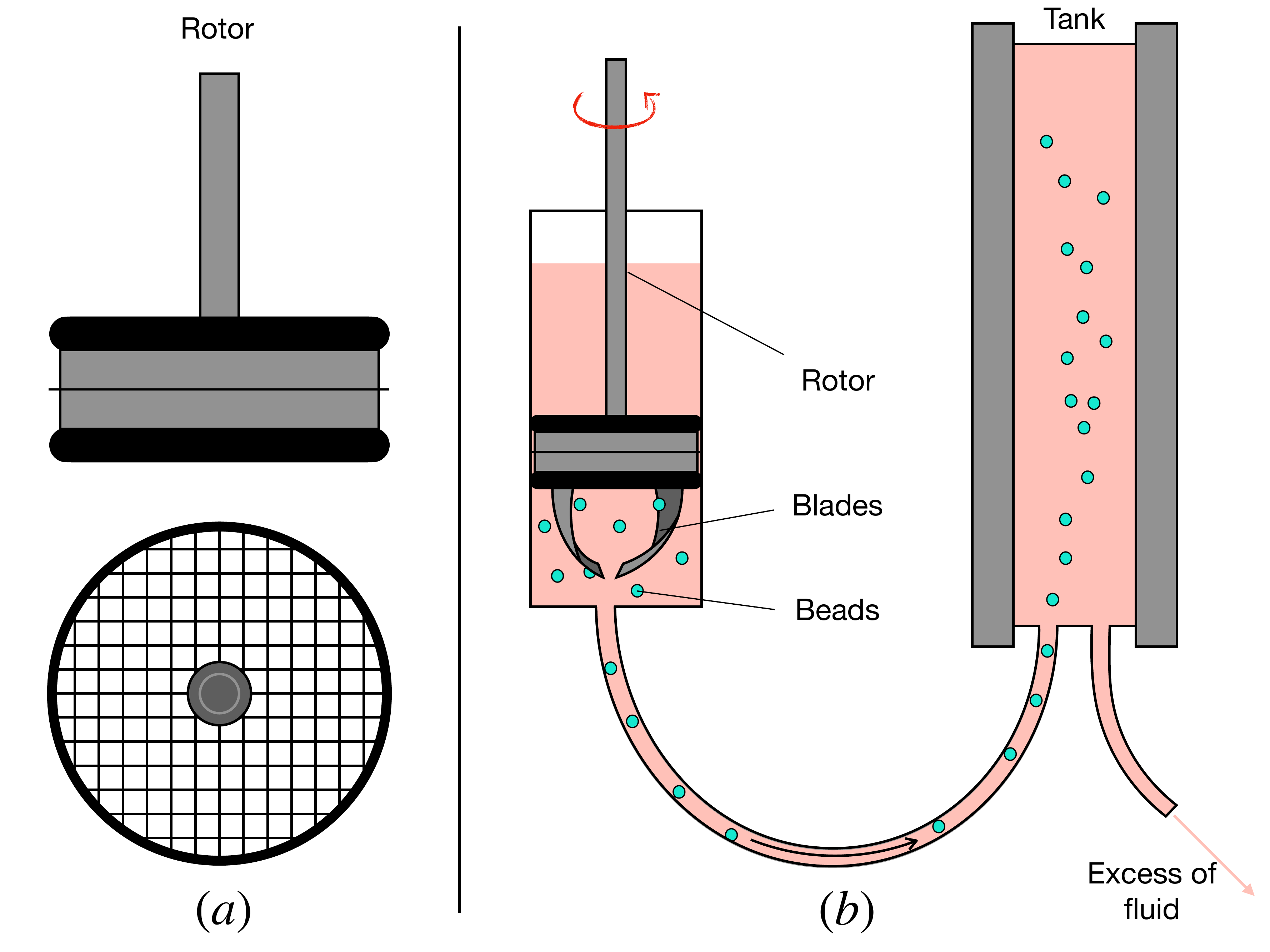}
	\caption{Stir-machine that enables the introduction of beads without air bubbles. $(a)$ The grid in the rotor has a mesh smaller than the beads diameter, that retains them in the lower part of the cylinder while letting the bubbles escape. $(b)$ Set up to introduce beads in the tank in the vertical position. The grid and the rotor are immersed in the fluid. Blades have also been fixed on the rotor, in order to keep beads in suspension, to avoid their settling against the grid and to allow bubbles detachment and escape through the grid.}
	\label{fig:stir}
\end{figure}

\section{Snapshots of the top and bottom deposits}
\indent Figure \ref{fig:stir} shows two pictures of the deposits studied: (a) the floating crust that forms at the surface, and (b) the cumulate that appears when some conditions enable its formation. Picture (a) has been taken at the end of an experiment, when all particles have settled at the surface. Note the egg-box-like shape of the floating lid, very similar to shape that can be observed in geomorphology and in experiments by \cite{Solomatov93a}. Concerning the basal cumulate (d), note that deposits form patterns like dunes. We must emphasize  that these dunes move and deform as the convection goes on. 

\begin{figure}[h]
	\centering
	\includegraphics[width=0.8\textwidth]{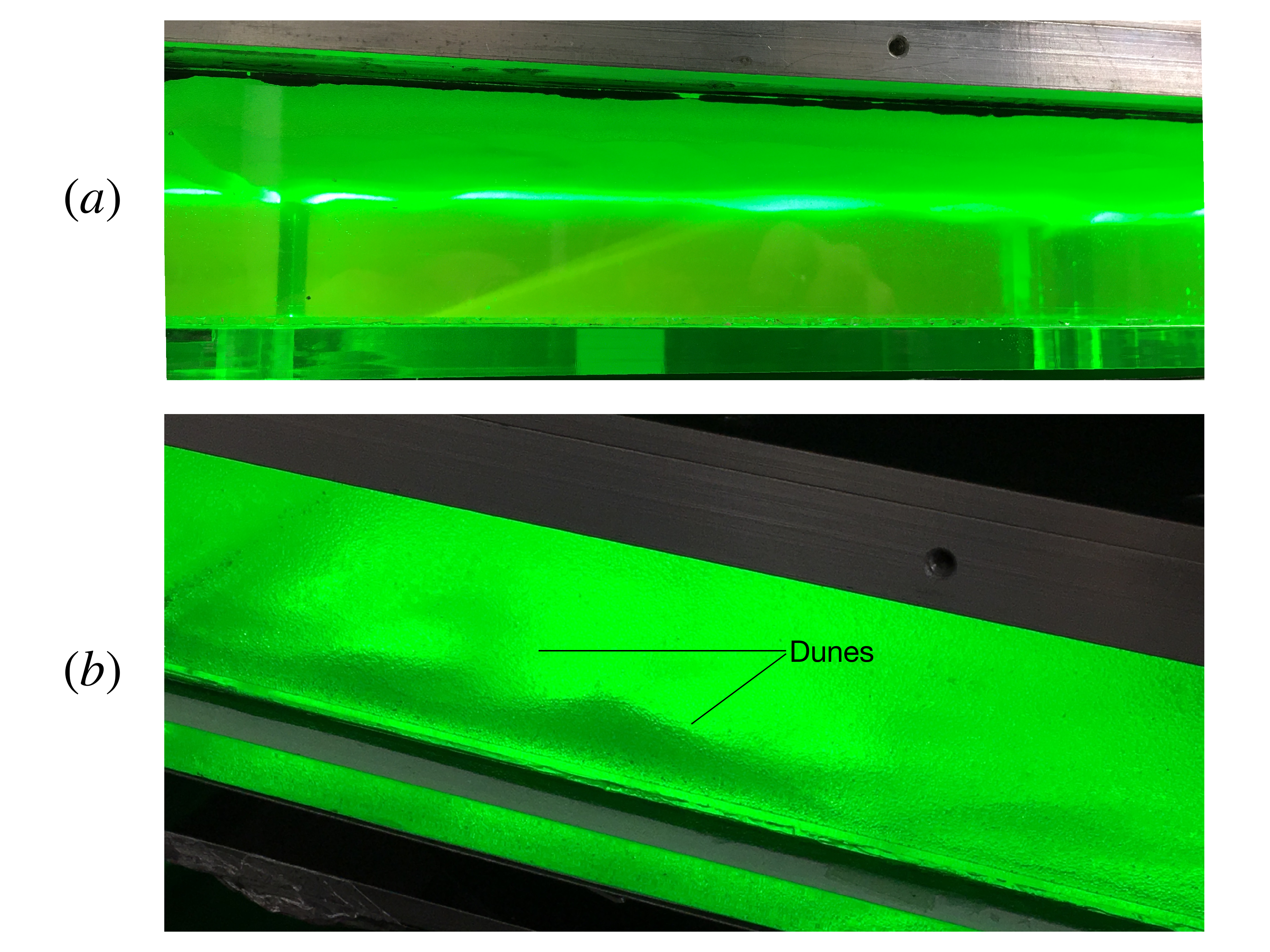}
	\caption{Different types of deposits that can form during experiments. $(a)$: picture of settled beads at room temperature. $(b)$: snapshot of the cumulate forming at the end of a hot experiment. Note that both deposits are dune-shaped.}
	\label{fig:stir}
\end{figure}

\section{Data for the scaling laws dealing with strain rates and velocities}
\indent Refractive index mismatch between the fluid and beads affected the velocity field as soon as the laser sheet scanned larger distances from the front wall and the concentration of particles became significant. We performed 8 experiments without any beads in order to determine accurately the scaling laws for the velocity and the strain rate. We measured the steady-state velocity field in order to calculate either the characteristic velocities (vertical and horizontal) and the characteristic strain rate (vertical and horizontal). All these values are evaluated by their root mean square (RMS) values calculated over the entire volume of the tank. The results are shown in Table \ref{tab:scaling}. These data were used to determine the scaling laws shown in section 4.2.2.

\begin{landscape}

\topskip0pt
\vspace*{\fill}

\begin{table}[htb!]
   \centering
   \begin{tabular}{| c |  c c c c c c c  |}
      \hline	
      Name 		& $Ra_H$ ($10^7$)	& $T_{s}$ ($^{o}C$) & $T_{bulk}$ ($^{o}C$)	& $U_x$	($10^{-4}$ m/s)		& $U_z$ ($10^{-4}$ m/s)		& $\dot \gamma_{xz}$ ($10^{-2}$/s) 	&	$\dot \gamma_{zx}$ ($10^{-2}$/s)		\\
      \hline
      IHB29\_1 	& 	0.28			& 22.2			&		26.0			& 2.64					&	2.64					&  4.21						&		4.19		\\
      IHB29\_2	& 	3.4			& 22.2			&		34.5			&7.96  					&	7.48					&  13.7						& 		14.8 		\\
      IHB29\_3 	&	7.9			& 22.3			& 		38.9			&6.26 					&	6.11 					& 10.3						& 		11.6		\\
      IHB30\_1    &	0.77			& 22.3			&		28.5			&3.72					&	3.64					& 5.77						&		6.00		 \\
      IHB30\_2	&  	1.8			& 22.4			&		31.4			&5.09					&	4.50					&  7.69						&		7.64		\\
      IHB30\_3 	&	12.2			& 25.6			&		43.1			&7.68					&	6.66					&	16.2						&	16.3 \\
      IHB31\_1	&	0.16			& 9.1				&		16.5			&2.29					&	1.76					& 4.28						&		3.66			\\
      IHB31\_2	&	13.3			& 29.4			&		46.6			&10.0					&	 8.83					& 18.6						& 19.7\\
	\hline
   \end{tabular}
   \caption{Experimental conditions (steady-state $Ra_{H}$ and surface and bulk temperatures $T_{s}$ and $T_{\mathrm{bulk}}$ respectively) and RMS values of the horizontal velocity ($U_{x}$), vertical velocity ($U_{z}$), horizontal strain rate ($\dot \gamma _{xz}$) and vertical strain rate ($\dot \gamma_{zx}$), obtained in homogeneous, beads-free experiments.}
   \label{tab:scaling}
\end{table}

\vspace*{\fill}

\end{landscape}

\bibliographystyle{unsrtnat}
\bibliography{bibliosm}